\documentclass{elsarticle}
\usepackage[utf8]{inputenc}
\usepackage[T1]{fontenc}

\usepackage{geometry}[leftmargin=1.5cm,rightmargin=1.5cm]

\usepackage{tensor}

\usepackage{pgfplots}
\pgfplotsset{compat=newest}

\usepackage{graphicx}
\usepackage{rotating}
\usepackage{pifont}
\usepackage{xcolor}
\usepackage{tabu}
\usepackage{multirow}
\usepackage{makecell}
\usepackage{lscape}
\usepackage{enumitem}
\usepackage[hidelinks]{hyperref}

\usepackage{isotope}

\usepackage{scalefnt}

\graphicspath{{./figures/}}

\usepackage{amsmath}
\usepackage{amssymb}
\usepackage{amsfonts}
\usepackage{bm}
\usepackage{xfrac}
\usepackage{braket}

\usepackage{lmodern}
\usepackage{microtype}

\usepackage{scalefnt}

\newcommand{\overlap}[2]{\langle #1 |#2 \rangle}

\newcommand{\cdotss}{\mathinner{\cdotp\mkern-3mu\cdotp\mkern-3mu\cdotp}}
\newcommand{\cdotsm}{\mathinner{\cdotp\mkern-2mu\cdotp\mkern-2mu\cdotp}}

\newcommand{\floor}[1]{\left\lfloor#1\right\rfloor}
\newcommand{\ceil}[1]{\left\lceil#1\right\rceil}
\newcommand{\bq}[1]{\left[#1\right]}
\renewcommand{\Re}{\mathrm{Re}}
\newcommand{\nuc}[2]{\tensor*[^{#1}]{\mathrm{#2}}{}}

\usepackage[normalem]{ulem}

\newcommand{\ud}{\mathrm{d}}

\newcommand{\ts}{\textsuperscript}

\newcommand{\inv}[1]{\frac{1}{#1}}

\newcommand{\bmbptuncon}{BMBPT$^{\circ}$}
\newcommand{\bmbptcon}{BMBPT$^{\bullet}$}
\newcommand{\bmbptap}{BMBPT$^{\ast}$}
\newcommand{\ciuncon}{BCI$^{\circ}$}
\newcommand{\cicon}{BCI$^{\bullet}$}
\newcommand{\ciap}{BCI$^{\ast}$}
\newcommand{\EX}{\text{BCI}}

\makeatletter
\def\ps@pprintTitle{%
	\let\@oddhead\@empty
	\let\@evenhead\@empty
	\def\@oddfoot{\footnotesize\itshape
		Accepted manuscript by Annals of Physics\hfill\today}%
	\let\@evenfoot\@oddfoot}
\makeatother

\newcommand*{\copyrightII}{%
	\raisebox{-0.3\baselineskip}{%
		\includegraphics[
		height=1.1\baselineskip,
		keepaspectratio,
		]{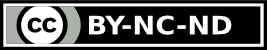}%
	}%
}

\tnotetext[]{\hspace*{-5mm}\href{http://creativecommons.org/licenses/by-nc-nd/4.0/}{\copyrightII} \ This manuscript version is made available under the  \href{http://creativecommons.org/licenses/by-nc-nd/4.0/}{\textit{\underline{CC-BY-NC-ND 4.0 license}}}}

\begin{document}

\title{Bogoliubov many-body perturbation theory under constraint}

\address[kul]{KU Leuven, Department of Physics and Astronomy, Instituut voor Kern- en Stralingsfysica,\\ 3001 Leuven, Belgium}
\address[cea]{IRFU, CEA, Universit\'e Paris-Saclay, 91191 Gif-sur-Yvette, France}
\address[mpik]{Max-Planck-Institut f\"ur Kernphysik, 69117 Heidelberg, Germany}
\address[ikp]{Institut f\"ur Kernphysik, Technische Universit\"at Darmstadt, 64289 Darmstadt, Germany}
\address[emmi]{ExtreMe Matter Institute EMMI, GSI Helmholtzzentrum f\"ur Schwerionenforschung GmbH, \\ 64291 Darmstadt, Germany}
\address[esnt]{ESNT, IRFU, CEA, Universit\'e Paris-Saclay, 91191 Gif-sur-Yvette, France}

\author[kul]{P. Demol\corref{cor1}}
\ead{pepijn.demol@kuleuven.be}
\cortext[cor1]{Corresponding author}

\author[cea]{M. Frosini}
\ead{mikael.frosini@cea.fr}

\author[mpik,ikp,emmi,esnt]{A. Tichai}
\ead{alexander.tichai@physik.tu-darmstadt.de}

\author[cea]{V. Som\`a}
\ead{vittorio.soma@cea.fr}

\author[kul,cea]{T. Duguet}
\ead{thomas.duguet@cea.fr}

\begin{abstract}

In order to solve the A-body Schr\"odinger equation both accurately and efficiently 
for open-shell nuclei, a novel many-body method coined as Bogoliubov many-body perturbation 
theory (BMBPT) was recently formalized and applied at low orders. Based on the breaking of $U(1)$ symmetry associated with particle-number conservation, this perturbation theory must operate under the constraint that the {\it average} number of particles is self-consistently adjusted at each perturbative order. The corresponding formalism is presently detailed with the goal to characterize the behaviour of the associated Taylor series. BMBPT is, thus, investigated numerically up to high orders at the price of restricting oneself to a small, i.e. schematic, portion of Fock space. While low-order results only differ by $2-3\,\%$ from those obtained via a configuration interaction (CI) diagonalization, the series is shown to eventually diverge. The application of a novel resummation method coined as eigenvector continuation further increases the accuracy when built from low-order BMBPT corrections and quickly converges towards the CI result when applied at higher orders. Furthermore, the numerically-costly self-consistent particle number adjustment procedure is shown to be safely bypassed via the use of a computationally cheap \textit{a posteriori} correction method. Eventually, 
the present work validates the fact that low order BMBPT calculations based on an \textit{a posteriori} (average) particle number correction deliver controlled results and demonstrates that they can be optimally complemented by the eigenvector continuation method to provide results with sub-percent accuracy. This approach is, thus, planned to become a workhorse for realistic \textit{ab initio} calculations of open-shell nuclei in the near future.

\end{abstract}


\maketitle

\section{Introduction} 

The long-term objective of the so-called \textit{ab initio} approach 
to atomic nuclei is to develop an accurate and universal description of low-energy nuclear systems from first principles. 
Such a viewpoint stipulates that the atomic nucleus can be appropriately modeled in terms of $A=N+Z$ structureless and strongly interacting neutrons and protons. 
In this context, the basic interactions between proton and neutron degrees of freedom emerge from the underlying gauge 
theory of interacting quarks and gluons, i.e., from quantum chromodynamics (QCD). 
As such, the \textit{ab initio} endeavour involves two steps:
\begin{enumerate}[label=\arabic*)]
	\item Modelling the elementary inter-nucleon interactions (ideally with an 
	uncertainty estimate);
	\item  Solving the $A$-body Schr\"odinger equation (ideally with an uncertainty 
	estimate);
\end{enumerate}
such that the output predictions to be confronted with experimental data are a 
convolution of these two components. 
Whenever the uncertainty associated with one of 
the two above components dominates, the distance to the 
data can be attributed to it, thus, leading to the necessity to 
improve on it. Eventually, the \emph{ab initio} approach offers a systematic path towards a 
universal theoretical framework to describe nuclear properties ranging from binding energies and charge radii to spectroscopic properties and electroweak transition probabilities.

In the past 15 years, \textit{ab initio} low-energy nuclear theory has made 
tremendous progress regarding points 1) and 2) above. 
First, the \textit{ab initio} approach has been systematically formulated within the 
frame of chiral effective field theory ($\chi$-EFT) \cite{EntemMach,EpelbaumHamer} in which 
quark and gluon, as well as heavy hadron, degrees of freedom are integrated out. The long- and mid-range parts of inter-nucleon interactions are  mediated by pions, the Goldstone bosons of the spontaneously broken chiral symmetry at low energy, and are complemented with contact interactions accounting for high-energy degrees of freedom that are not  explicitly incorporated\footnote{Complementary EFTs have been designed to deal with phenomena characterized by even lower resolution scales~\cite{Hammer:2019poc}.}.  The low-energy constants (LEC's) of the EFT Lagrangian are typically fixed by reproducing a selected set of few-body data.  As such, $\chi$-EFT yields (i) a sound connection to QCD, (ii) a clear hierarchy of  the importance of two-nucleon (NN) interactions, three-nucleon (3N) interactions, \ldots, (iii) a consistent construction of other, e.g. electroweak, operators, (iv) a mean to estimate uncertainties due to the truncation employed in the systematic construction of the operators and (v) a systematic way to improve on the description if necessary. In addition to their construction within the frame of $\chi$-EFT, another key development relates to the use of similarity renormalization group (SRG) transformations to ''soften'' nuclear Hamiltonians and make them more amenable to many-body calculations~\cite{Jurgenson2009, Bogner2010}. The unitary SRG evolution constitutes a \emph{pre-diagonalization} of the operator in momentum space, thus suppressing the coupling between high- and low-momentum modes. As a result, many-body applications discussed below based on SRG-evolved operators have shown highly improved model-space convergence, thus facilitating studies of mid-mass nuclei.

As for point 2), the continuous improvement of methods formulated in the 1980s to solve the $A$-nucleon Schr\"odinger equation, as well as the development of novel ones, have allowed the computation of many more nuclear observables from first principles. As a first step, essentially exact solutions were typically provided by large scale diagonalization methods such as the no-core shell model (NCSM)~\cite{NavratilQuaglioni, BarrettNavratil} and by Green’s function Monte Carlo (GFMC) techniques \cite{PudlinerPandharipande, PieperWiringa}. However, due to the exponentially scaling cost with respect to basis/system size, these approaches are typically limited\footnote{Complemented with importance truncation techniques~\cite{Roth2009,Stumpf2016}, NCSM calculations can nowadays typically reach $A\approx 24$. } to light nuclei with mass number $A \lesssim  12$. In this context, a breakthrough occurred about 15 years ago to access heavier doubly closed-shell nuclei, i.e., nuclei whose neutron and proton numbers are such that the highest occupied single-nucleon shells are fully filled in a simple mean-field description. This breakthrough was made possible thanks to the development and application of non-perturbative methods whose numerical cost scale polynomially with system size. Examples are coupled cluster (CC)~\cite{DeanHjorth, KowalskiDean, HagenPapenbrock, BinderLanghammer, PiecuchGour}, in-medium similarity renormalization group (IM-SRG)~\cite{TsukiyamaBogner, HergertBogner2013, MorrisParzuchowski, HergertBogner2016} and self-consistent Green’s function (SCGF)~\cite{PhysRevC.65.064313, PhysRevC.68.014311, DickhoffBarbieri, CipolloneBarbieri, Soma20c} methods. 
In particular, while SCGF advanced within the field of nuclear physics, CC was successfully transferred back from quantum chemistry where it has been intensively developed over the last four decades to describe molecular properties from first principles.
These methods have allowed one to access a variety of observables in a few tens of doubly closed-(sub)shell nuclei with $10 \lesssim A \lesssim 100$. 

In principle, the combined use of $\chi$-EFT Hamiltonians and sophisticated methods to solve the A-body Schr\"odinger equation provides a universal framework with high predictive power. 
Most remarkably, benchmark calculations in closed-(sub)shell oxygen isotopes ($A \sim 20$) have demonstrated the consistency among the various many-body techniques and proved that their current level of implementation delivers ground-state observables with an uncertainty better than $2-3\,\%$~\cite{Hebeler15}.
Following this achievement, many-body calculations also acquired the role of diagnostic tools and have been used to test qualities and deficiencies of input Hamiltonians across the whole range of medium-mass nuclei~\cite{Soma:2013xha, LapouxSoma, TichaiArthuis}. 
At present, a strong effort is devoted to a better understanding of the shortcomings of existing $\chi$-EFT Hamiltonians with the ambition to improve on the accuracy 
of \textit{ab initio} calculations in the future \cite{NNLOsat,Soma20,Huther:2019ont}. 

Many-body theories accessing mid-mass nuclei typically expand the exact ground-state wavefunction 
around a reference Slater determinant and can meaningfully access doubly 
closed-shell systems. However, they are not suited to \textit{open-shell} systems that 
constitute the large majority of nuclei. This limitation is due to the fact that the 
ground-state wavefunction of open-shell nuclei is not dominated by a single Slater 
determinant such that the Hartree-Fock (HF) approximation cannot yield an 
appropriate reference point for the expansion. Alternatives have been developed 
to overcome this cutting-edge difficulty. A first option relies on the derivation of effective 
valence-space Hamiltonians that are subsequently diagonalized to access the spectrum 
of the target nucleus. While initially developed within a perturbative scheme 
\cite{Holt2013}, valence-space interactions have recently been formulated within non-perturbative NCSM~\cite{Dikmen:2015tla}, CC 
\cite{Soma:2013xha} and IM-SRG \cite{BognerHergert} frameworks. Still, the 
dimension of the associated configuration space grows exponentially when moving away 
from shell closures, which makes it eventually difficult to use such methods beyond $A 
\sim 100$. A second flavour of many-body methods applicable to open-shell nuclei are equation-of-motion 
(EOM) techniques, where one starts from the solution obtained for 
a closed-shell nucleus and describes neighbouring nuclei via the action of 
particle-attachment or particle-removal operators. While this has been extensively 
used in CC theory~\cite{PiecuchGour}, current implementations are restricted to the 
attachment/removal of at most two particles~\cite{Jansen:2012ey}, which prohibits its use through large degenerate single-particle shells. 

Generally speaking, the restriction to a single Slater-determinant reference state is too limiting to design a meaningful expansion  method directly in open-shell nuclei due to the degeneracy with respect to elementary particle-hole excitations. The use  of more general reference states must be contemplated \cite{RolikSzabados, SurjanRolik} to lift the degeneracy and tackle, from the outset, strong static correlations associated with it. The first option in this direction relies on reference states mixing a set of appropriately chosen Slater determinants. Those multi-configurational reference states can, for 
example, be obtained from a prior NCSM calculation in small model spaces or under the form of a
particle-number-projected Hartree-Fock-Bogoliubov (PHFB) state. Such reference 
states have been successfully employed in the multi-reference extension of IM-SRG (MR-IMSRG) 
\cite{HergertBogner2014,GebrerufaelVobig} or within a perturbative framework 
yielding multi-configurational perturbation theory (MCPT) \cite{TichaiMCPT}. 

With the objective to maintain a strict polynomial cost with basis/system size and keep the intrinsic simplicity of  single-reference expansion methods, another option relies on
reference states breaking one or several symmetries, i.e. states that do not carry an eigenvalue of the Casimir operator of a given symmetry of the Hamiltonian as a good quantum number. In semi-magic nuclei, $U(1)$ global-gauge symmetry associated with particle-number conservation is allowed to break in order to address Cooper pair's instability and handle nuclear superfluidity. This leads to expanding the exact 
solution of the $A$-body Schr\"odinger equation around a Bogoliubov 
reference state that reduces to a Slater determinant in closed-shell systems. In 
doubly open-shell nuclei, $SU(2)$ rotational symmetry associated with 
angular-momentum  conservation must also be allowed to break, thus, leading to the use of a 
spatially-deformed reference state. The above considerations have led to 
the design of non-perturbative Gorkov self-consistent Green's function (GSCGF)
\cite{SomaDuguet2011, Soma14a} and Bogoliubov coupled cluster (BCC) 
\cite{SignoracciDuguet} methods that generalize standard SCGF and CC to open-shell systems. Restricting oneself to a perturbative method, this rational has led to the design of Bogoliubov many-body perturbation theory (BMBPT)~\cite{DuguetSignoracci} that is the focus of the present paper\footnote{The breaking of symmetries is only {\it emergent} in a finite quantum system but not actually realized~\cite{ui83a,yannouleas07a,Papenbrock:2013cra}, i.e. it simply constitutes a (tremendously useful!) artefact of an approximate  calculation such that symmetries of the nuclear states must eventually be restored. With this objective in mind, even more general formalisms coined as particle-number 
and/or angular-momentum projected BCC (PBCC) and BMBPT (PBMBPT) have been recently 
formulated \cite{DuguetSignoracci, Duguet2014}. While the application of PBCC to the 
schematic solvable Richardson Hamiltonian is very encouraging \cite{QiuHenderson}, 
the full-fledged implementation of PBMBPT and PBCC to nuclei is still awaiting.}.

Focusing so far on singly open-shell nuclei, the formal and numerical developments 
of symmetry-breaking many-body methods have led to unprecedented achievements in 
the past years.
A notable example are the first systematic \textit{ab initio} calculations along 
complete chains of oxygen, calcium and nickel isotopes \cite{LapouxSoma, Soma:2013xha, Soma20} via GSCGF 
theory.
This method has then been applied in the neighbourhood of singly-magic calcium, e.g. in argon~\cite{Mougeot20}, potassium~\cite{PapugaBissell, Sun20} and other chains up to chromium~\cite{Soma20b}.
These calculations have contributed to the characterization of the too 
limited quality of existing chiral EFT Hamiltonians in mid-mass nuclei few years back~\cite{Soma:2013xha,PapugaBissell,LapouxSoma,Duguet:2016wwr}. Recently, GSCGF was employed to perform the first exploratory calculation of Sn and Xe isotopes \cite{arthuis2020ab}, extending the current range of applicability of \textit{ab initio} calculations to $A \sim 140$. A few years ago, BMBPT was implemented up to third order and shown to provide an accurate description of medium-mass ground-state energies at a significantly lower computational cost than GSCGF, BCC or MR-IMSRG theory \cite{TichaiArthuis}. This makes BMBPT an extremely useful candidate to perform large survey calculations, to systematically test next generations of chiral EFT Hamiltonians~\cite{Tichai:2020dna} and to make the future extension to even more challenging doubly open-shell nuclei simpler than in other \textit{ab initio} frameworks.

The use of a perturbation theory relies on the hope that the associated Taylor series converges or possesses asymptotic properties that justify the use of the first few orders as a meaningful estimate of the full series. This question has been addressed in Refs. \cite{RothPade} and \cite{TichaiMBPT} 
for standard MBPT appropriate to doubly closed-shell nuclei. Despite softening the interaction via an SRG transformation~\cite{Bogner2007} to tame down its ultra-violet source of non-perturbative character~\cite{Bogner:2005,Ramanan:2007,Hoppe:2017}, MBPT was shown~\cite{RothPade} to typically diverge in small model spaces for $\nuc{16}{O}$. The algebraic Padé resummation method was successfully invoked to recover physical quantities from the diverging series. Reference \cite{TichaiMBPT} further revealed that using the variationally optimised HF Slater determinant as a reference state (i.e. using the M\o{}ller-Plesset scheme) in combination with an SRG-softened  interaction provides a convergent MBPT series. These results are consistent with what was found earlier on in quantum chemistry~\cite{leininger00a}.

The objective of the present paper is to extend this study to (singly) open-shell nuclei studied through BMBPT. Strong  static correlations of infrared origin are "regularized" via the breaking of $U(1)$ symmetry such that a meaningful expansion can \textit{at least} be defined on top of the reference state. While low orders have indeed been shown to provide sound results~\cite{TichaiArthuis}, the behaviour of the associated Taylor series remains to be characterized more thoroughly by pushing BMBPT to high orders. 

Furthermore, it happens that the breaking of, e.g. $U(1)$, symmetry has profound consequences on the characteristics of the series produced through the perturbative expansion. While the reference state and the perturbatively corrected states are not \textit{eigenstates} of the Casimir operator of the symmetry group, one wishes to impose that the targeted eigenvalue is at least reproduced \textit{on average}\footnote{This feature is not restricted to perturbation theory and must equally be imposed in non-perturbative schemes based on symmetry-breaking reference states such as GSCGF or BCC.}. Breaking $U(1)$ symmetry thus implies that states at play are not eigenstates of the particle-number operator $A$ but must carry at least the physical number of particle on average\footnote{It must be clear from the outset that this constraint on the average particle number is \textit{not} identical to the actual restoration of good particle number accomplished by PBMBPT or PBCC~\cite{DuguetSignoracci, QiuHenderson} on top of BMBPT or BCC.}. In this context, the difficulty relates to the fact that the average particle number typically changes at each perturbative order. In Ref.~\cite{TichaiArthuis}, low orders were addressed in such a way that the shift of the average particle number occurring at each order was accounted for by an unsubstantiated \textit{a  posteriori} correction.  A more robust formalism was suggested by the authors in which the average particle number is actually adjusted to the correct value at each perturbative order. This idea leads to a new type of unexplored perturbative sequence characterized by the following two features:
\begin{enumerate}[label=\arabic*)]
	\item Even though the exact solution obtained as the limit of the sequence must 
	lie within the Hilbert space associated with $A$-particle systems, the sequence 
	itself is not restricted to that Hilbert space and spans the entire Fock space.
	\item The expansion involves in fact two coupled sequences associated with the 
	energy and the average particle number such that the latter is constrained to match
	the targeted physical value at each working order. The coupling between the two 
	sequences and the need to deliver the physical particle number on average at each order makes the approach intrinsically iterative and at 
	variance with standard MBPTs. 
\end{enumerate} 
 Eventually, one is led to considering a new type of expansion coined as many-body perturbation theory under constraint\footnote{A similar feature arises within the frame of MBPT-based orbital-dependent density functional theory, i.e. the perturbative expansion must be constrained to yield no corrections to the local density such that the reference Kohn-Sham state displays the same local density as the fully correlated/corrected state at each working order, e.g. see Ref.~\cite{PTUCinDFT}.}. The presently introduced constrained version of BMBPT is denoted as \bmbptcon~while the unconstrained form is indicated as \bmbptuncon. The third variant of BMBPT employed in Ref.~\cite{TichaiArthuis} makes use of an \textit{a posteriori} correction and is denoted as \bmbptap. 
 
In this context, the objective of the present study is to investigate the following, yet unexplored, aspects of BMBPT (or rather \bmbptcon): 
\begin{enumerate}[label=\arabic*)]
	\item What is the high-order behaviour of the perturbative expansion under constraint?
	\item In absence of convergence, does this behaviour authorize the use of standard or novel resummation methods delivering the correct result?
	\item If so, do low orders provide a fair approximation of the resummed series? 
	\item Is the \textit{a posteriori} correction employed in Ref.~\cite{TichaiArthuis} justified such that the iterative and costly character of \bmbptcon~can be entirely bypassed in actual applications via the use of \bmbptap?
\end{enumerate} 

To address these various points, and contrary to its original derivation based on a time-dependent formalism~\cite{DuguetSignoracci,Pierre}, \bmbptcon~is presently introduced on the basis of a more traditional time-independent approach. In this context, \bmbptcon~is easily formulated via a recursive scheme from which corrections up to high, e.g.,  20\ts{th} or 30\ts{th}, order can be efficiently computed. Still, doing so in numerical applications requires to limit oneself to a small, i.e. schematic, portion of Fock space such that a severe truncation on the set of eigenstates of the unperturbed Hamiltonian must be considered. It is the price to pay to be able to investigate the series up to high orders and one hopes that the truncation effects do not invalidate the general conclusions reached in this way.

The document is organized as follows. In Sec.~\ref{sec:ME}, basic equations of the many-body problem are stated and notations are introduced. Section~\ref{sec:BF} introduces the basic ingredients of the Bogoliubov framework. In Sec.~\ref{sec:BMBPT}, the \bmbptcon~formalism is developed and compared to standard unconstrained MBPT, i.e. \bmbptuncon. After introducing the Taylor series associated with strict perturbation theory, the well-known Pad\'e  resummation scheme and the recently proposed eigenvector continuation (EC) technique~\cite{Frame,Frame:2019jsw} are introduced. In Sec.~\ref{sec:res}, results from calculations performed within a small model-space are displayed and analyzed. The specificities of \bmbptuncon, \bmbptcon~ and \bmbptap, as well as of the resummation methods built on them, are probed and validated against exact diagonalization. Lastly, conclusions and perspectives are provided in Sec.~\ref{sec:con}. 

\section{Master equations}
\label{sec:ME}

\subsection{Eigenvalue equations}
\noindent
\textit{Ab initio} nuclear structure calculations aim at solving the 
time-independent many-body Schr\"odinger equation 
\begin{equation}
H |\Psi_n^\text{A}\rangle = \text{E}_n^{\text{A}} |\Psi_n^{\text{A}} \rangle  
\hspace{2mm}, 
\end{equation}
where $H$ is the Hamiltonian defined from elementary inter-nucleon interactions, while $|\Psi_n^\text{A}\rangle$ and $\text{E}_n^\text{A}$ denote its A-body eigenstates and eigenenergies, respectively.  As testified by the carried quantum number\footnote{The reader is advised not to be confused between the operator $A$ (math style) and its eigenvalue A (roman style) used throughout this work.} A, the Hamiltonian commutes with the particle-number operator, i.e. $[H,A]=0$.

Due to the allowed breaking of particle-number symmetry later on, the actual problem of interest consists in fact of explicitly considering the two coupled eigenvalue equations
\begin{subequations}
	\begin{align}
	A |\Psi_n^\text{A}\rangle &= \text{A} |\Psi_n^\text{A} \rangle  
	\, ,\\ 
	\Omega |\Psi_n^\text{A}\rangle &= \mathcal{E}_n^\text{A} |\Psi_n^\text{A} 
	\rangle  \, \label{eq:pot} ,
	\end{align}
\end{subequations}
where the grand potential operator is defined through
\begin{equation} \label{eq:grandpot}
\Omega \equiv H - \lambda A \, ,
\end{equation}
with $\lambda$ the chemical potential. The eigenstates $|\Psi_n^\text{A}\rangle$ of $\Omega$ are the same as those of $H$ and carry eigenvalues $\mathcal{E}_n^\text{A}\equiv \text{E}_n^{\text{A}} - \lambda \text{A}$. Generically, the aim is thus to solve an eigenvalue problem 
\begin{equation}\label{eq:eigO}
O |\Psi_n^\text{A}\rangle = \text{O}_n^\text{A} |\Psi_n^\text{A} \rangle  
\hspace{2mm}, 
\end{equation}
for an operator $O$ such that $[H,O]=0$, while invoking a constraint. Solving the coupled eigenvalue equations in a perturbative way leads to an iterative formalism developed in Sec.~\ref{sec:BMBPT}.

\subsection{Evaluation of observables}\label{sec:obs}

Given a quantum state, there exist essentially two general ways to evaluate an observable, i.e. the so-called \textit{projective} and \textit{expectation value} methods. Historically, a projective approach is typically used to compute the energy in the context of MBPT and CC. In contrast, the expectation value method is traditionally used in SCGF. The two methods coincide in the exact limit, i.e., they deliver the eigenvalue of interest when the considered quantum state is indeed an eigenstate of the corresponding operator. Due to the approximate solving of the A-body problem, however, the two approaches yield different results in practice.

\subsubsection{Projective approach}
\label{sec:proj}

The projective measure of an observable $O$ in the quantum state $|\Psi_n^\text{A} 
	\rangle$ is defined as 
\begin{equation}\label{eq:projsym}
\mathcal{O}_n^\text{A} \equiv \dfrac{1}{2} \left( \dfrac{\langle \Phi| O 
	|\Psi_n^\text{A} 
	\rangle}{\langle \Phi| \Psi_n^\text{A} \rangle} + \dfrac{\langle 
	\Psi_n^\text{A}| O |\Phi 
	\rangle}{\langle \Psi_n^\text{A}|\Phi \rangle} \right)  = \Re \left\{ \dfrac{\langle \Phi| O 
	|\Psi_n^\text{A} 
	\rangle}{\langle \Phi| \Psi_n^\text{A} \rangle}\right\}\ \ ,
\end{equation}
where $|\Phi \rangle$ is an appropriate reference state such that $\langle \Phi| \Psi_n^\text{A} \rangle \neq 0$. This expression is manifestly real whenever $O$ is self-adjoint. If $|\Psi_n^\text{A} \rangle$ is an eigenstate of $O$, each term in the parenthesis is actually real and delivers the corresponding eigenvalue $\mathcal{O}_n^\text{A} = \mathrm O_n^\text{A}$. However, if the exact eigenstate is approximated, the first term might become complex\footnote{Each term is real at any order of BMBPT if the operator $O$ is the one driving the perturbative expansion~\cite{Pierre}, i.e. for $O=\Omega$. In standard MBPT based on a Slater  determinant reference state, each term is also real for $O=H$ at each truncation order (in this case, the particle number is conserved and there is no point considering $A$ or $\Omega$). In all other cases, the symmetrized expression of Eq.~\eqref{eq:projsym} must be used.}, thus, the use of a symmetric definition in Eq.~\eqref{eq:projsym}. 

Similarly, the dispersion of $O$ in $|\Psi_n^\text{A} \rangle$ is defined through
\begin{subequations} \label{eq:varpr}
	\begin{align}
	\Delta \mathcal{O}_n^\text{A} &\equiv \Re \left\{ \dfrac{\langle \Phi| (O - 
		\mathcal{O}_n^\text{A})^2  |\Psi_n^\text{A} \rangle}{\langle \Phi| 
		\Psi_n^\text{A} \rangle} 
	\right\}\\
	&=  \Re \left\{ \dfrac{\langle \Phi| O^2 |\Psi_n^\text{A} \rangle - 2  
		\mathcal{O}_n^\text{A} \cdot \langle \Phi| O |\Psi_n^\text{A} \rangle 
		+ 
		(\mathcal{O}_n^\text{A})^2 
		\cdot \langle \Phi|\Psi_n^\text{A} \rangle }{\langle \Phi| 
		\Psi_n^\text{A} \rangle} 
	    \right\}\\
	&=  (\mathcal{O}^2)_n^\text{A} - (\mathcal{O}_n^\text{A})^2  \ .
	\end{align}
\end{subequations}
If $|\Psi_n^\text{A} \rangle$ is an eigenstate of $O$ with eigenvalue $\mathrm O_n^\text{A}$, it is also an eigenstate of $O^2$ 
with eigenvalue $(\mathrm O_n^\text{A})^2$ such that $\Delta \mathcal{O}_n^\text{A} = 0$ as expected. 
The inverse statement is not true, i.e., $\Delta \mathcal{O}_n^\text{A}$ can vanish without $|\Psi_n^\text{A} \rangle$ being an eigenstate of $O$. 

\subsubsection{Expectation value approach} \label{sec:exp}

The expectation value measure of $O$ in $|\Psi_n^\text{A} 
	\rangle$ is straightforwardly defined as 
\begin{equation}\label{eq:expval}
\langle O \rangle_n^\text{A} \equiv \dfrac{\langle \Psi_n^\text{A}|O| 
	\Psi_n^\text{A} \rangle 
}{\langle \Psi_n^\text{A}|\Psi_n^\text{A} \rangle} \ \ ,
\end{equation}
which is manifestly real if $O$ is self-adjoint.
If $| \Psi_n^\text{A} \rangle$ is an eigenstate of $O$, one has $\langle O \rangle_n^\text{A} = \text{O}_n^\text{A}$. 
Consistently, the variance is given by 
\begin{subequations} \label{eq:varex}
	\begin{align}
	\Delta \langle O \rangle_n^\text{A} &\equiv  \dfrac{\langle 
		\Psi_n^\text{A}|(O - \langle O 
		\rangle_n^\text{A} )^2 | \Psi_n^\text{A} \rangle 
	}{\langle \Psi_n^\text{A}|\Psi_n^\text{A} \rangle}\\
	&= \langle O^2\rangle_n^\text{A} - (\langle O \rangle_n^\text{A})^2 \ .
	\end{align}
\end{subequations}
Here, $| \Psi_n^\text{A} \rangle$ is an eigenstate of $O$ with eigenvalue $\langle O \rangle_n^\text{A}$ if and only if $\Delta \langle O \rangle_n^\text{A} = 0$. While potentially interesting, the expectation value approach is not investigated in the present document such that only results obtained from the projective measure are reported below. 

\section{Bogoliubov framework\label{sec:BF}}

The novelty of single-reference BMBPT is to employ a particle-number breaking Bogoliubov reference state as  a way to handle open-shell nuclei in a controlled fashion. The present section introduces the basics of Bogoliubov algebra, Bogoliubov vacua and the associated Wick's theorem.

\subsection{Bogoliubov algebra}

Quasi-particle annihilation and creation operators $\{\beta^\dagger_k, \beta_k \}$ are related to a set of ordinary particle operators $\mathcal{B}_1 \equiv\{c_p^{\dagger}, c_p\}$ making up a basis of the one-body Hilbert space $\mathcal{H}_1$ via a Bogoliubov transformation \cite{RingSchuck}
\begin{subequations}\label{eq:bt}
	\begin{align}
	\beta_k &\equiv \sum_p U_{pk}^* c_p +  V_{pk}^* c_p^{\dagger} \ , \\
	\beta_k^{\dagger} &\equiv \sum_p U_{pk} c_p^{\dagger} +  V_{pk} c_p \ .
	\end{align}	 
\end{subequations}
This linear transformation can be written in matrix form as
\begin{equation}
\begin{pmatrix}
\beta \\
\beta^{\dagger}
\end{pmatrix}
= W^{\dagger}
\begin{pmatrix}
c \\
c^{\dagger}
\end{pmatrix}
\ ,
\end{equation}
where
\begin{equation}
W \equiv
\begin{pmatrix}
U & V^{\ast} \\
V &  U^{\ast}
\end{pmatrix}
\ . \label{bogomatrix}
\end{equation}
Enforcing that both sets of fermionic operators obey anti-commutation rules
\begin{subequations}
	\label{anticom}
	\begin{align}
	\{c_{p},c_{q}\} &= 0 \ \ , &  
	\{c^{\dagger}_{p},c^{\dagger}_{q}\} &= 0\ \ , &  
	\{c_{p},c^{\dagger}_{q}\} &= \delta_{pq} \ \ \ ; \\
	\{\beta_{k_1},\beta_{k_2}\} &= 0  \ \ , & 
	\{\beta^{\dagger}_{k_1},\beta^{\dagger}_{k_2}\} &= 0 \ \ , &  
	\{\beta_{k_1},\beta^{\dagger}_{k_2}\} &= \delta_{k_1k_2} \ , \label{eq:anticom}
	\end{align}
\end{subequations}
translates into the fact that $W$ is unitary, allowing one to invert the Bogoliubov transformation. 

Given the set of quasi-particle operators, the Bogoliubov many-body state $|\Phi \rangle $ is introduced as their vacuum satisfying $\beta_k |\Phi \rangle = 0 $ for all $k$, which defines it up to a phase. Because quasi-particle operators mix particle creation and annihilation operators,  $|\Phi \rangle $ is not an eigenstate of the particle number operator\footnote{At this point, other symmetries of the Hamiltonian are not necessarily broken. For example, in the present work, quasi-particle operators carry orbital angular 
momentum quantum number $l$, total angular momentum $j$ and magnetic quantum number 
$m$ as good quantum numbers as a testimony of rotational symmetry. In addition, a label $t$ represents the isospin projection characteristic of a neutron or a proton. Eventually, a principal quantum label $n$ is necessary to fully specify each quasi-particle state.  The index $k$ is therefore in one-to-one correspondence with the set of quantum numbers $(n, l, j, m, t)$.}. Still, one typically requires that the Bogoliubov state carries the physical number of particles\footnote{In fact, the proton number $Z$ and neutron number $N$ are broken independently and in a practical application one needs to constrain them separately to the right value by using two different Lagrange parameters.} on average, i.e.
\begin{equation}
\langle \Phi | A | \Phi \rangle = \text{A} \, .
\end{equation}

In practice, one needs to specify how the Bogoliubov vacuum is effectively obtained, i.e, the set of quasi-particle operators $\{\beta_{k}^{\dagger}, \beta_{k}\}$, or equivalently the matrices $U$ and $V$ making up the Bogoliubov transformation matrix $W$, must be characterized. In the present study,  $|\Phi \rangle $ is taken to be the solution of the variational problem formulated within the manifold of Bogoliubov states, i.e. the $U$ and $V$ matrices solve the Hartree-Fock-Bogoliubov eigenvalue equation~\cite{RingSchuck} generalizing the HF one. This standard mean-field variational problem is briefly recalled in \ref{ap:HFB}.

\subsection{Operator representation \label{sec:operator}}

Given $\mathcal{B}_1 \equiv\{c_p^{\dagger}, c_p\}$, a generic operator $O$ commuting with $A$ and summing a one-body, a two-body\ldots, up to a A-body contribution is written in second-quantized form according to 
\begin{equation}
O = \dfrac{1}{(1!)^2} \sum_{pq} o^{1N}_{pq}
c^{\dagger}_{p} c_{q} + \dfrac{1}{(2!)^2}\sum_{pqrs}
 \bar  o^{2N}_{pqrs}
c^{\dagger}_{p}c^{\dagger}_{q}c_{s}c_{r}\\
 +\dfrac{1}{(3!)^2}\sum_{pqrstu} \bar o^{3N}_{pqrstu}
c^{\dagger}_{p}c^{\dagger}_{q}c^{\dagger}_{r}c_{u}c_{t}c_{s}
+ \cdots \hspace{2mm},
\end{equation}
where matrix elements $ o^{1N}_{pq}$, $\bar o^{2N}_{pqrs}$, $\bar o^{3N}_{pqrstu}$ are fully anti-symmetric with respect to permutations within the groups of indices associated with creation or annihilation operators. Using Wick's theorem \cite{Wick} with respect to the Bogoliubov vacuum, the operator can be rewritten as a sum of normal-ordered products of quasi-particle creation and annihilation operators
\begingroup
\allowdisplaybreaks
\begin{subequations} \label{eq:operator}
	\begin{align} 
	O \equiv & O^{[0]} + O^{[2]} + O^{[4]} + \cdots\\
	\equiv &O^{00} + ( O^{20} + O^{11} + O^{02}) + ( O^{40} + O^{31} + O^{22} + 
	O^{13} + O^{04}) + \cdots\\
	= & O^{00}  \nonumber  \\ \nonumber
	& + \left( \dfrac{1}{2!}\sum_{k_1k_2} O^{20}_{k_1k_2} \beta^{\dagger}_{k_1} 
	\beta^{\dagger}_{k_2} + \dfrac{1}{1!}\sum_{k_1k_2} O^{11}_{k_1k_2} 
	\beta^{\dagger}_{k_1} \beta_{k_2} +	\dfrac{1}{2!}\sum_{k_1k_2} 
	O^{02}_{k_1k_2} 
	\beta_{k_2}\beta_{k_1}\right) \\\nonumber
	& + \Bigg(  \dfrac{1}{4!}\sum_{k_1k_2k_3k_4} O^{40}_{k_1k_2k_3k_4} 
	\beta^{\dagger}_{k_1} \beta^{\dagger}_{k_2} \beta^{\dagger}_{k_3} 
	\beta^{\dagger}_{k_4} + \dfrac{1}{3! \ 1!} \sum_{k_1k_2k_3k_4} 
	O^{31}_{k_1k_2k_3k_4}\beta^{\dagger}_{k_1} \beta^{\dagger}_{k_2} 
	\beta^{\dagger}_{k_3} \beta_{k_4} \\ \nonumber
	& + \dfrac{1}{(2!)^2} \sum_{k_1k_2k_3k_4} O^{22}_{k_1k_2k_3k_4} 
	\beta^{\dagger}_{k_1} \beta^{\dagger}_{k_2} \beta_{k_4} \beta_{k_3} + 
	\dfrac{1}{1! \ 3!} \sum_{k_1k_2k_3k_4} O^{13}_{k_1k_2k_3k_4} 
	\beta^{\dagger}_{k_1}\beta_{k_4} \beta_{k_3} \beta_{k_2}\\
	& + \dfrac{1}{4!} \sum_{k_1k_2k_3k_4} O^{04}_{k_1k_2k_3k_4}	\beta_{k_4} 
	\beta_{k_3} \beta_{k_2} \beta_{k_1} \Bigg) + \cdots\\
	&=\sum_{k=0,2,4,6 \cdots} \ \sum_{\substack{i=0 \\ i+j=k}}^k
	\ \dfrac{1}{i! \ j!} \ \sum_{\substack{k_1,k_2, \cdots,k_i \\ k_{i+1}, 
	k_{i+2},\cdots,k_{i+j}}} 
	 O^{ij}_{k_1\cdots k_i k_{i+1} \cdots k_{i+j}} 
	\beta^{\dagger}_{k_1} \cdots \beta^{\dagger}_{k_i} \beta_{k_{i+j}} \cdots 
	\beta_{k_{i+1}} \ \ .
	\end{align}
\end{subequations}

\endgroup
\noindent Here $O^{ij}_{k_1\cdots k_i k_{i+1} \cdots k_{i+j}}$ is totally anti-symmetric with respect to permutations of indices belonging either to the subset of annihilation operators or to the subset of creation operators. 
The expressions of $O^{ij}_{k_1\cdots k_i k_{i+1} \cdots k_{i+j}}$ up to $O^{[6]}$ 
as a function of the $U$ and $V$ Bogoliubov matrices as well as of the matrix elements $\bar o^{kN}_{p_1p_2 \cdots 
p_k}$ with $k \leq 3$, can be found in Refs. \cite{SignoracciDuguet,NOKB}. 

For example, the above can be straightforwardly applied to the nuclear Hamiltonian $H \equiv T + V + W$ initially defined as
\begin{equation}\label{eq:Hamil}
H \equiv \dfrac{1}{(1!)^2} \sum_{pq} t_{pq}
c^{\dagger}_{p} c_{q} + \dfrac{1}{(2!)^2}\sum_{pqrs}
\bar  v_{pqrs}
c^{\dagger}_{p}c^{\dagger}_{q}c_{s}c_{r} +\dfrac{1}{(3!)^2}\sum_{pqrstu} \bar 
w_{pqrstu}
c^{\dagger}_{p}c^{\dagger}_{q}c^{\dagger}_{r}c_{u}c_{t}c_{s}\hspace{2mm},
\end{equation}
where $T$ denotes the kinetic energy and where anti-symmetric matrix elements $ \bar  v_{pqrs}$ and $ \bar w_{pqrstu}$ of 2N and 3N interactions are employed. The same is true for the particle number operator   
\begin{equation}
A \equiv \sum_{p} 
c^{\dagger}_{p} c_{p}  \hspace{2mm} .
\end{equation}

\section{Bogoliubov many-body perturbation theory\label{sec:BMBPT}}

Bogoliubov many-body perturbation theory expands exact eigenstates $\ket{\Psi_n^{\text A}}$ of $H$ around a Bogoliubov reference  state breaking particle-number symmetry. It presents the tremendous advantage that static, i.e.  pairing, correlations at play in (singly) open-shell nuclei are already largely accounted for by the reference state. In doing so, the degeneracy of Slater determinants with respect to particle-hole excitations is lifted from the outset\footnote{To describe doubly 
open-shell nuclei, the spherical symmetry of $\ket{\Phi}$ associated with rotational invariance must be relaxed.}, thus offering the possibility to meaningfully expand the exact eigenstates on top of it.

As already mentioned, a constraint on the average particle number needs to be imposed at each order 
in the BMBPT expansion. Fixing the average particle number in the reference state to a targeted value, e.g. the physical one, it is shifted by the perturbative corrections\footnote{The 	  
counting of the orders in this text is different from the one used in Ref. 
\cite{TichaiArthuis}. Here, the unperturbed solution is labelled as order zero	
while Ref. \cite{TichaiArthuis} starts at order one. Consequently, the counting presently employed is shifted by one order compared to the standard one, e.g. the usual (B)MBPT(2) is presently coined as first order.}. As a result, one must in fact envision to enforce that the \textit{sum} of the contributions equates the physical value at each  working order. In the following section, MBPT under constraint is thus developed within the framework of BMBPT and is denoted as \bmbptcon\footnote{The formal derivation is in fact general and can therefore be adapted to any auxiliary constraint associated with an 
operator $O$ commuting with $H$.}.

\subsection{Perturbative expansion under constraint}
\label{MBPTsetup}

In order to monitor the average particle number, a Lagrange term is added to the Hamiltonian\footnote{The chemical potential $\lambda$ is fixed such that $\mathcal{E}_0^{\text{A}_0}$ for the targeted particle number $\text{A}_0$ is the lowest value of all $\mathcal{E}_n^{\text{A}}$ over Fock space, i.e. it penalises systems with larger number of particles such that $\mathcal{E}_0^{\text{A}_0} \leq \mathcal{E}_n^{\text A}$ for all $\text A \geq \text A _0$ while maintaining at the same time that $\mathcal{E}_0^{\text A_0} \leq \mathcal{E}_n^{\text A}$ for all $\text A \leq \text A _0$.  This is practically achievable only if $\text{E}_0^\text{A}$ is strictly convex in the neighbourhood of $\text A _0$, which is generally but not always true for atomic  nuclei~\cite{DuguetSignoracci}.}, thus leading to the introduction of the grand potential in Eq.~\eqref{eq:grandpot}. To set up the perturbation theory, the driving operator $\Omega$ is partitioned according to
\begin{equation}\label{eq:partPindep}
\Omega  =  \Omega_0 +  \Omega_1 \ ,
\end{equation}
where $\Omega_0$ ($\Omega_1$) denotes the unperturbed (residual) part. The key feature of BMBPT is to permit both $\Omega_0$ and  $\Omega_1$ to break $U(1)$ global gauge symmetry, i.e. to authorize that $[\Omega_0, A] \neq 0$ and  $[\Omega_1, A] \neq 0$. The operator $\Omega_0$ is chosen such that its eigenbasis can be constructed exactly, thus providing the set of unperturbed states among which the state carrying the lowest (non-degenerate) eigenvalue is nothing but the reference state  $|\Phi \rangle $. 

The particle-number shift induced by the perturbative corrections is anticipated from the outset and accounted for by adjusting the average particle number carried by the reference state accordingly. This is done\footnote{As discussed in Sec.~\ref{sec:PNA}, this calls for an iterative procedure.} by adapting $\lambda$, which actually corresponds to a redefinition of $\Omega$. This procedure must be conducted at each working order\footnote{The precise meaning of $P$ will be defined retrospectively through Eq.~\eqref{eq:stateOP}.} $P$, leading in fact to the use of an {\it order-dependent} grand potential operator
\begin{equation} \label{eq:omegaPdep}
\Omega_P  \equiv H - \lambda_P A \ , 
\end{equation} 
where the label $P$ indicates the perturbative order at which the constraint is meant to be imposed.  This feature constitutes a key novelty of \bmbptcon~whose consequences are discussed at length in the following. Notice that $\Omega_P$ commutes with $H$ for each value of $P$, so that the eigenbasis of $\Omega_P$ is independent of $P$ and can be taken to be the same as the eigenbasis of $H$. On the other hand, the associated eigenvalues $\mathcal{E}_{n P}^\text{A}\equiv E_{n}^\text{A} - \lambda_P \text{A}$ are now $P$-dependent. The practical implementation of the particle-number adjustment procedure is the subject of Sec.~\ref{sec:PNA}. 

\subsubsection{Order-dependent partitioning}

Due to the $P$-dependence introduced in Eq.~\eqref{eq:omegaPdep}, the partitioning 
of Eq.~\eqref{eq:partPindep} must be rewritten as
\begin{equation}
\Omega_P  =  \Omega_{0 P} +  \Omega_{1 P} \ , \label{Ppartitioning}
\end{equation} 
such that the reference state $| \Phi_P \rangle$ itself depends on $P$. More explicitly, the partitioning is stipulated under the form
\begin{subequations}
	\begin{align}
	\Omega_{0 P} & \equiv \Omega_P^{00} + \bar{\Omega}_P^{11} \ , \\ 
	\Omega_{1 P} & \equiv \Omega_P^{20} + \breve{{\Omega}}_P^{11} + 
	\Omega_P^{02} 
	+   
	\Omega_P^{40} + \Omega_P^{31} + \Omega_P^{22} + \Omega_P^{13} + 
	\Omega_P^{04} + \cdots 
	\ \ ,
	\end{align}
\end{subequations}
where the normal-ordered representation with respect to  $| \Phi_P \rangle$ defined through Eq.\eqref{eq:operator} is employed, with $\breve{{\Omega}}^{11}_P \equiv \Omega^{11}_P - 
\bar{\Omega}^{11}_P $. The unperturbed part $\Omega_{0 P}$ contains the zero-body 
operator (number) 
\begin{equation}
\Omega_{P}^{00}  =  \dfrac{\langle \Phi_P | \Omega_P | \Phi_P\rangle}{\langle \Phi_P | \Phi_P\rangle} 
\end{equation} 
and a diagonal one-body operator chosen to take 
the form 
\begin{equation} \label{diagonebodypiece}
\bar{\Omega}_P^{11} \equiv \sum_k E_{k P} \beta^{\dagger}_{k P} \beta_{k P}\ \ ,
\end{equation}
where $E_{k P}$ denotes a set of positive quasi-particle energies. In practice, the HFB solution is utilized as the Bogoliubov vacuum and constitutes the zero-order approximation to the perturbative sequence. This choice of reference state corresponds to the M\o{}ller-Plesset scheme in standard MBPT. Working in this scheme implies that $\Omega^{02}_P = \Omega^{20}_P = \breve{\Omega}^{11}_P = 0$ and that the quasi-particle energies $E_{k P}$ are taken as the solutions of the HFB eigenvalue equation (Eq.~\eqref{eq:HFBeigen}). Still, BMBPT equations are presently derived for a generic Bogoliubov vacuum such that the M\o{}ller-Plesset  scheme is easily obtained by setting $\Omega^{02}_P, \Omega^{20}_P$ and  $\breve{{\Omega}}^{11}_P $ to zero at the end. 

Acting with all possible strings\footnote{The quasi-particle vacuum $|\Phi_P \rangle$ itself is included in the set as a zero quasi-particle excitation.}\ts{,}\footnote{Targeting even-even nuclei, as is done in the present document, all basis states carrying an even number of quasi-particle excitations span the accessible part of Fock space. Therefore it is enough to only include Bogoliubov states of even quasi-particle rank into the basis.} of quasi-particle creation operators on the vacuum generates the many-body states
\begin{equation}\label{eq:unperstates}
|\Phi_P^{k_1k_2 \cdots}\rangle \equiv \beta_{k_1  P}^\dag \beta_{k_2  P}^\dag \cdots
|\Phi_P\rangle \ ,
\end{equation}
where the number of quasi-particle excitations characterizes the rank of the state. It is easy to verify that the eigenbasis of the unperturbed operator $\Omega_{0 P}$ is given by
\begin{subequations}\label{eq:zeroo}
	\begin{align}
	\Omega_{0 P} |\Phi_P \rangle &= \Omega^{00}_P |\Phi_P \rangle 
	\label{eq:zeroo1}\\ 
	\Omega_{0 P} |\Phi_P^{k_1k_2 \cdots} \rangle &= ( \Omega_P^{00} +
	{E_{k_1k_2\cdots P}}) |\Phi_P^{k_1k_2 \cdots}\rangle \hspace{2mm} 
	\label{eq:zeroo2} ,
	\end{align}
\end{subequations}
where
\begin{equation}
E_{k_1k_2\cdots P} \equiv E_{k_1  P} + E_{k_2  P} + \cdots \ . \label{NqpExcit}
\end{equation}

\subsubsection{Wave-function expansion\label{sec:wfexp}}

Concentrating on a generic order $P$, an auxiliary operator is further introduced
\begin{equation}\label{eq:split}
\Omega_P (x) \equiv \Omega_{0 P} + x \ \Omega_{1 P} \ \ ,
\end{equation} 
where $x \in [0,1]$ denotes the expansion parameter, such that 
\begin{subequations}
	\begin{align}
	{\Omega_P} (0) &=  \Omega_{0 P}\, ,  \\
	{\Omega_P} (1) &=   \Omega_P \, .
	\end{align}
\end{subequations}
Eigenvalues and eigenvectors of $\Omega_P (x)$ are defined through
\begin{equation}\label{eq:eigx}
\Omega_P (x) |\Psi_{n P} (x) \rangle = \tilde{\mathcal{E}}_{n P} (x) |\Psi_{n P} 
(x)\rangle \ 
\ ,
\end{equation}
such that
\begin{subequations}
	\begin{align}
	\lim_{x \to 1} |\Psi_{n P} (x) \rangle &=  |\Psi_{n P}^\text{A} \rangle = 
	\ket{\Psi_n^{\text A}} \ ,  \\
	\lim_{x \to 1} \tilde{\mathcal{E}}_{n P} (x) &= \mathcal{E}_{n P}^\text{A} \ ,
	\end{align}
\end{subequations}
deliver the eigenvector and eigenvalue of $\Omega_P$, respectively. One notices that 
$|\Psi_{n P} (x) \rangle$ and $\tilde{\mathcal{E}}_{n P} (x)$ do not carry the superscript A in general. Indeed, exact eigenstates of $\Omega_P(x)$ do themselves break particle-number symmetry given that $[\Omega_P(x),A]\neq0$ except for $x=1$, i.e. only in the limit $x \to 1$ is $U(1)$ global gauge symmetry satisfied.  

Next, a power series expansion of $|\Psi_{n P} (x) \rangle$ and  $\tilde{\mathcal{E}}_{n P} (x)$ in terms of $x$ is formulated
\begin{subequations}\label{eq:exp}
	\begin{align}
	&|\Psi_{n P} (x)\rangle \equiv |\Phi_{n P}^{(0)}\rangle +  x \ 
	|\Phi_{n P}^{(1)}\rangle +  
	x^2 \  
	|\Phi_{n P}^{(2)}\rangle + ... = |\Phi_{n P}^{(0)}\rangle +  \sum _{p\geq1}  x^p 
	\ 
	|\Phi_{n P}^{(p)}\rangle \ \label{eq:expstate} \ , \\
	&\tilde{\mathcal{E}}_{n P} (x)\equiv \tilde{\mathcal{E}}_{n P}^{(0)} + x \ 
	\tilde{\mathcal{E}}_{n P}^{(1)} +  x^2 \ \tilde{\mathcal{E}}_{n P}^{(2)} + 
	\cdots = 
	\tilde{\mathcal{E}}_{n P}^{(0)} +\sum _{p\geq1}  x^p \ 
	\tilde{\mathcal{E}}_{n P}^{(p)} \	
	\ ,
	\end{align}
\end{subequations}
where the upper index $(p)$ labels each coefficient in the power series. Intermediate normalization\footnote{Using this convention, $|\Psi_{n P} (x)\rangle$ is not normalized as soon as $x\neq0$.} is invoked
\begin{equation}
\langle\Phi^{(0)}_{n P}|\Psi_{n P}(x)\rangle = 1 \hspace{8mm}\forall \ x  \ ,
\end{equation}
which is ensured by
\begin{subequations}
	\label{eq:int_norm}
	\begin{align}
	\langle\Phi^{(0)}_{n P}|\Phi^{(0)}_{n P}\rangle &= 1 \ ,\\
	\langle\Phi^{(0)}_{n P}|\Phi_{n P}^{(q)}\rangle &= 0 \ , \hspace{8mm} \forall \ 
	q\geq 1\ .
	\end{align}
\end{subequations}

Evaluating Eq.~\eqref{eq:eigx} at $x=0$ yields
\begin{equation}
\Omega_{0 P} |\Phi_{n P}^{(0)}\rangle = \tilde{\mathcal{E}}_{n P}^{(0)} 
|\Phi_{n P}^{(0)}\rangle 
\  \ , 
\end{equation}
which is nothing but the eigenvalue equation for the unperturbed grand potential 
$\Omega_{0 P}$. Using Eq.~\eqref{eq:zeroo}, one can further identify
\begin{subequations}
	\begin{align}
	&|\Psi_{n P} (0)\rangle = |\Phi_{n P}^{(0)}\rangle \equiv  
	|\Phi_{P}^{k_1k_2\cdots} 
	\rangle 
	\label{eq:unperstate} \, , \\
	&\tilde{\mathcal{E}}_{n P}(0) = \tilde{\mathcal{E}}_{n P}^{(0)} \equiv 
	\Omega^{00}_{P} + E_{k_1k_2\cdots P}	\label{eq:unperE}\ ,
	\end{align}
\end{subequations}
which for the ground state reduces to the Bogoliubov reference state of the order-$P$ calculation, i.e. $|\Phi_{0 P}^{(0)}\rangle \equiv  |\Phi_{P} \rangle$ and $\tilde{\mathcal{E}}_{0 P}^{(0)}=\Omega^{00}_{P}$. 

The actual \emph{$P$-order} perturbative approximation of $|\Psi_{n P}(x) \rangle$ is obtained from Eq.~\eqref{eq:expstate} by truncating the power series at order $P$
\begin{equation}\label{eq:stateOP}
|\Psi_{n P}^{[P]} (x) \rangle\equiv \sum_{p=0}^P x^p |\Phi_{n P}^{(p)}\rangle \ ,
\end{equation}
where the summation index $P$ eventually {\it defines} the perturbative order used by anticipation in Eqs.~\eqref{eq:omegaPdep}-\eqref{eq:unperE}. The procedure is summarized through the set of equations
\begin{subequations} \label{eq:setaprox}
	\begin{align}
	| \Psi_{n \, 0}(x) \rangle &= 
	\overbrace{|\Phi_{n \, 0}^{(0)}\rangle}^{|\Psi_{n \, 0}^{[0]}(x)\rangle}
	+ x|\Phi_{n \, 0}^{(1)}\rangle
	+ x^2|\Phi_{n \, 0}^{(2)} \rangle+ x^3|\Phi_{n \, 0}^{(3)}\rangle +
	\cdots + x^p|\Phi_{n \, 0}^{(P)}\rangle + \cdots \ , \label{eq:setaproxA} \\
	| \Psi_{n \, 1}(x)\rangle &= \overbrace{|\Phi_{n \, 1}^{(0)}\rangle+ 
		x|\Phi_{n \, 1}^{(1)}\rangle}^{|\Psi_{n \, 1}^{[1]}(x)\rangle}
	+ x^2|\Phi_{n \, 1}^{(2)} \rangle + x^3|\Phi_{n \, 1}^{(3)}\rangle +
	\cdots + x^p|\Phi_{n \, 1}^{(P)} \rangle+ \cdots \ ,  \\
	| \Psi_{n \, 2}(x)\rangle &= \overbrace{|\Phi_{n \, 2}^{(0)}\rangle + 
		x|\Phi_{n \, 2}^{(1)}\rangle + 
		x^2|\Phi_{n \, 2}^{(2)}\rangle}^{|\Psi_{n \, 2}^{[2]}(x)\rangle} 
	+ x^3|\Phi_{n \, 2}^{(3)} \rangle + \cdots + x^p|\Phi_{n \, 2}^{(P)} \rangle+ \cdots \ 
	, \\
	& \ \, \vdots& \notag \\
	| \Psi_{n P}(x)\rangle &= \underbrace{|\Phi_{n P}^{(0)}\rangle + 
		x|\Phi_{n P}^{(1)}\rangle  + x^2|\Phi_{n P}^{(2)} \rangle+ 
		x^3|\Phi_{n P}^{(3)}\rangle +
		\cdots + x^p|\Phi_{n P}^{(P)}\rangle}_{|\Psi_{n P}^{[P]}(x)\rangle} + \cdots 
	\ ,  \label{eq:setaproxD}
	\end{align}%
\end{subequations}
such that the sequence of states $\big\lbrace  \ket{\Psi^{[P]}_{n P} (1)} \ 
\big | \  P =0,..., \infty \big\rbrace $ defines the successive 
approximations to the eigenstate $\ket{\Psi_n^A}$ of $H$ and are all required to  fulfil the  auxiliary constraint
\begin{equation}
\Re	\left \{ \bra{\Phi^{(0)}_{n P}} A \ket{\Psi^{[P]}_{n P} (1)} \right\}  = \text{A} \ 
.
\end{equation} 

In contrast to traditional MBPT based on a single partitioning leading to one Taylor series, \bmbptcon~generates a sequence of approximations, each of which  refers to a different partitioning and a different Taylor series, i.e. the approximations generated at each order are \textit{not} the successive partial sums associated with a single Taylor series.

If the constraint is relaxed, this framework reduces to the naive unconstrained BMBPT, i.e. \bmbptuncon, for which the definition of the driving operator, its splitting  and the Taylor series are independent of the order at which one eventually wishes to work. Consequently, there is only one power series expansion of the wave-function, i.e. Eqs.~\eqref{eq:setaproxA}-\eqref{eq:setaproxD} reduce to a single equation. In this case, the successive approximations to the wave-function are nothing but the consecutive partial sums of this single series. 

It is a compelling question whether or not one can eventually bypass the need to explicitly enforce the constrained  and design an efficient scheme in which \bmbptuncon~is complemented with an \textit{a posteriori} correction. If so, the chemical potential would typically be chosen such that the reference state carries the physical particle number on average, which corresponds to setting the subscript $P$ to $0$ independently of the actual order $[P]$ at which one wishes to operate. This question of great practical interest will be addressed in Sec.~\ref{aposterioriconstraintmethod}.

\subsubsection{Recursive scheme \label{sec:rec}}

In this section, a recursive scheme for the determination of the state corrections $|\Phi_{n P}^{(p)}\rangle $ is introduced as derived in Refs.~\cite{RothPade,TichaiMBPT}. Applying this scheme in a small model space allows to utilize BMBPT up to high orders. For notational convenience, the index $P$ characterizing the explicit $P$-dependence of the entire expansion scheme is dropped in the remainder of the paper. By default, the reader should keep in mind that $\Omega$, its partitioning, the associated unperturbed basis, the Taylor series etc. are actually $P$-dependent. 

Substituting Eqs.~\eqref{eq:split} and~\eqref{eq:exp} into Eq.~\eqref{eq:eigx} gives
\begin{equation} 
(\Omega_0 +  x \ \Omega_1) \left[|\Phi_n^{(0)}\rangle + \sum_{p\ge1}  x^p \ 
|\Phi_n^{(p)} \rangle\right] = \left[ \tilde{\mathcal{E}}_n^{(0)} +\sum 
_{p\geq1} x^p \ \tilde{\mathcal{E}}_n^{(p)} \right] \cdot 
\left[|\Phi_n^{(0)}\rangle 	+ \sum_{p\ge1} x^p \ |\Phi_n^{(p)} \rangle \right] \ ,
\end{equation}
such that grouping the terms proportional to $x^p$ leads to
\begin{equation}
\tilde{\mathcal{E}}_n^{(0)} |\Phi_n^{(0)} \rangle + \sum_{p\ge1}  x ^ p
\left[ \Omega_0 |\Phi_n^{(p)}\rangle  +
{\Omega}_1|\Phi_n^{(p-1)}\rangle\right] = \tilde{\mathcal{E}}_n^{(0)}
|\Phi_n^{(0)}\rangle + \sum_{p\ge1} x^p \left[\sum_{0\le j\le p}
\tilde{\mathcal{E}}_n^{(j)} |\Phi_n^{(p-j)}\rangle\right].
\label{eq:grouped}
\end{equation}
Left multiplying Eq.~\eqref{eq:grouped} with $\langle \Phi_n^{(0)}|$ and using 
intermediate normalization (Eq.~\eqref{eq:int_norm}) yields
\begin{equation}
\sum_{p\ge1}  x^p \langle\Phi_n^{(0)} | \Omega_1 |
\Phi_n^{(p-1)}\rangle = \sum_{p\ge1}  x^p \tilde{\mathcal{E}}_n^{(p)},
\end{equation}
which allows one to identify
\begin{equation}\label{eq:Ep}
\tilde{\mathcal{E}}_n^{(p)} = \langle\Phi_n^{(0)} |\Omega_1 | \Phi_n^{(p-1)}\rangle\, .
\end{equation}
Left multiplying Eq.~\eqref{eq:grouped} with $\langle\Phi_m^{(0)}|$, $m\neq
n$, further gives
\begin{equation}
\sum_{p\ge1}  x^p \left[\tilde{\mathcal{E}}_m^{(0)} \langle \Phi_m^{(0)} |
\Phi_n^{(p)} \rangle + \langle \Phi_m^{(0)} | \Omega_1 |
\Phi_n^{(p-1)} \rangle\right] = \sum_{p\ge1} x^p\left[\sum_{0\le j\le p} 
\tilde{\mathcal{E}}_n^{(j)} \langle \Phi_m^{(0)} |\Phi_n^{(p-j)}\rangle\right],
\end{equation}
such that matching the terms proportional to $x^p$ provides the relation
\begin{equation}\label{eq:Enp}
\left(\tilde{\mathcal{E}}_n^{(0)} - \tilde{\mathcal{E}}_m^{(0)}\right) \langle 
\Phi_m^{(0)} |\Phi_n ^{(p)} \rangle    =     \langle \Phi_m^{(0)}| \Omega_1 | 
\Phi_n^{(p-1)}\rangle  - \sum_{1\le j\le p} \tilde{\mathcal{E}}_n^{(j)} \langle 
\Phi_m^{(0)} |\Phi_n^{(p-j)}\rangle.
\end{equation}

Introducing the coefficients
\begin{equation}\label{eq:Cmnp}
C_{mn}^{(p)} \equiv \langle \Phi_m^{(0)} | \Phi_n ^{(p)} \rangle = 
\inv {\tilde{\mathcal{E}}_n^{(0)} - \tilde{\mathcal{E}}_m^{(0)}} 
\left[ \langle \Phi_m^{(0)}| \Omega_1 | \Phi_n^{(p-1)}\rangle
- \sum_{1\le j\le p} \tilde{\mathcal{E}}_n^{(j)} \langle \Phi_m^{(0)} |
\Phi_n^{(p-j)}\rangle \right]
\end{equation}
allows one to expand $ |\Phi_n^{(p)} \rangle$ on the unperturbed basis 
$\{|\Phi_m^{(0)} \rangle\}$
\begin{equation}\label{eq:psinp}
| \Phi_n^{(p)}\rangle = \sum_m C_{mn}^{(p)} |\Phi_m^{(0)} \rangle,
\end{equation}
such that Eq.~\eqref{eq:Ep} becomes
\begin{equation}\label{eq:Enp_rec}
\tilde{\mathcal{E}}_n^{(p)} = \sum_m \langle \Phi_n^{(0)} | \Omega_1 | \Phi_m^{(0)} 
\rangle C_{mn}^{(p-1)}.
\end{equation}
Inserting Eq.~\eqref{eq:psinp} into Eq.~\eqref{eq:Cmnp} eventually provides 
\begin{equation} \label{eq:cnmp_rec}
C_{mn}^{(p)} = \inv {\tilde{\mathcal{E}}_n^{(0)} - \tilde{\mathcal{E}}_m^{(0)}} 
\left[ \sum_q \langle \Phi_m^{(0)} | \Omega_1 |
\Phi_q^{(0)}\rangle C_{qn}^{(p-1)}
- \sum_{1\le j\le p} \tilde{\mathcal{E}}_n^{(j)} C_{mn}^{(p-j)}
\right].
\end{equation}
Equation~\eqref{eq:cnmp_rec} permits to compute the wave-function coefficients recursively\footnote{When a symmetry-conserving Slater determinant is employed as reference state,
the recursive scheme introduced above reduces to the one discussed in Ref. 
\cite{RothPade,TichaiMBPT}.} through a matrix-vector multiplication at each new order involving the matrix of $\Omega_1$ expressed in the unperturbed  basis. The necessary initial conditions at $p=0$ are extracted from Eqs.~\eqref{eq:int_norm} and~\eqref{eq:unperE} such that
\begin{subequations} 
	\begin{eqnarray}
	\tilde{\mathcal E}_n^{(0)}  &=& \Omega^{00} + E_{k_1k_2\cdots} \, , \\
	C_{mn}^{(0)}        &=& \delta_{mn} \, .
	\end{eqnarray}
\end{subequations}
As discussed in \ref{ap:subspace_state}, the subspace of Fock space $\mathcal F$ contributing to $|
\Phi_n^{(p)}\rangle$ can be identified by unfolding the recursive relation~\eqref{eq:cnmp_rec}. This characterization is of importance given that a truncation over the visited subspace is eventually performed in the numerical applications, i.e. $\Omega_1$ is represented only in a subspace of $\mathcal F$ when  building the matrix used to perform the repeated matrix-vector multiplications. The computed coefficients are, thus, complete only up to a certain perturbative order $P$, i.e. a growing number of terms are discarded when going to higher orders. One hopes that the general conclusions drawn out of the behaviour of the expansion up to high orders are however not affected.

\subsubsection{Matrix elements}

Up to this point, the fact that the unperturbed states $|\Phi_{n}^{(0)} \rangle$
are Bogoliubov states has not been explicitly exploited. The perturbative expansion is formally general and does not depend on the details of the partitioning employed. Eventually though, the working equations delivering $C^{(p)}_{mn}$, $\tilde{\mathcal E}^{(p)}_{n}$ and an observable $O$ (see below) are expressed in terms of the matrix elements of $ \Omega_{1}$ and $O$ in the unperturbed basis $|\Phi_{n}^{(0)} \rangle$ defined through Eqs.~\eqref{Ppartitioning}-\eqref{NqpExcit}. This feature is detailed in \ref{ap:ME}.

\subsection{Observable expansion \label{sec:obsexp}} 

The sequence of successive approximations $\big  \lbrace \ket{\Psi^{[P]}_{n} (1)}  ;  P = 0,..., \infty \big \rbrace $ to the eigenstate $\ket{\Psi_n^{\text{A}}} $ of $\Omega$ enables one to perturbatively calculate any observable $O$.  
Using the projective measure (Eq.~\eqref{eq:projsym}), the observable associated to $|\Psi_{n}^{[P]}(x) \rangle$ reads as
\begin{equation}\label{eq:projOP}
\mathcal{O}_{n}^{[P]}(x) \equiv
\Re\left\lbrace
\dfrac{\langle \Phi_{n}^{(0)}| O |\Psi_{n}^{[P]}(x) \rangle}
{\langle \Phi_{n}^{(0)}| \Psi_{n}^{[P]} (x)\rangle}
\right\rbrace =
\Re \left\lbrace \langle \Phi_{n}^{(0)}| O |\Psi_{n}^{[P]}(x)\rangle \right\rbrace
\ , 
\end{equation}
which is such that\footnote{This limit procedure has to be performed with care and in the same 
order as stated in Eq.~\eqref{eq:limit}.}
\begin{equation}\label{eq:limit}
\lim_{P \to \infty}\lim_{x \to 1} \mathcal{O}_{n}^{[P]} (x) =
\mathcal{O}_n^\text{A} \ .
\end{equation}
Substituting Eq.~\eqref{eq:stateOP} into Eq.~\eqref{eq:projOP} yields	
\begin{subequations}\label{eq:projOPex}
	\begin{align}
	\mathcal{O}_{n}^{[P]} (x) &= \Re\left\lbrace
	\sum _{p=0}^P x^p \, \langle \Phi_{n}^{(0)}| O |\Phi_n^{(p)}\rangle
	\right\rbrace\\
	&= \Re\left\lbrace
	\sum _{p=0}^P \sum_{m} x^p \ \langle \Phi_{n}^{(0)}| O 
	|\Phi_{m}^{(0)}\rangle C_{mn}^{(p)} 
	\right\rbrace\  \ , 
	\end{align}
\end{subequations}
where Eq.~\eqref{eq:psinp} was used. Therefore 
$\mathcal{O}_{n}^{[P]} (x)$ is a Taylor series in $x$ truncated at order $P$ such that $\big\lbrace 
\mathcal{O}_{n}^{[P]} (1) \ , \ P=0,\cdots,\infty 
\big\rbrace $ defines the sequence of successive approximations to O$^A_n$. Each term in the sequence originates from a different, i.e. order-dependent, Taylor series.

Considering the variance in its projective form
\begin{equation}\label{eq:projvarOP}
\Delta\mathcal{O}_{n}^{[P]} (x) \equiv (\mathcal{O}^2)_{n}^{[P]}(x) - 
(\mathcal{O}_{n}^{[P]} (x))^2 
\ ,
\end{equation}
and inserting Eq.~\eqref{eq:projOPex} leads to
\begin{equation}\label{eq:projvarOPex}
\Delta\mathcal{O}_{n}^{[P]}(x) = \Re\left\lbrace
\sum_{p=0}^P x^p
\sum_{m} \langle \Phi_{n}^{(0)}| O^2 |\Phi_{m}^{(0)}\rangle C_{mn}^{(p)} 
- \left(  \sum_{p=0}^P x^p \sum_{m}
\langle \Phi_{n}^{(0)}| O	|\Phi_{n}^{(0)}\rangle C_{mn}^{(p)}\right) ^2
\right\rbrace\ .
\end{equation}
Equation~\eqref{eq:projOPex} is applied to $O\equiv H, A$, and $\Omega$ to generate $E_{n}^{[P]} (x)$, $\mathcal{A}_{n}^{[P]} (x)$ and $\mathcal{E}_{n}^{[P]} (x)$, respectively. The same is done for the particle-number variance $\Delta \mathcal{A}_{n}^{[P]} (x)$ through Eq.~\eqref{eq:projvarOPex}.

Eventually, the subspace of $\mathcal F$ contributing to $\mathcal{O}_{n}^{[P]} (1)$ and $\Delta\mathcal{O}_{n}^{[P]} (1)$ at each order $P$ is investigated in \ref{ap:subspace_obs}.

\subsection{Resummation methods \label{sec:resum}}
\label{resummationmethods}

While the use of SRG-transformed Hamiltonians and of symmetry-breaking reference states tame down ultraviolet and infrared sources of non-perturbative behaviour~\cite{Tichai:2020dna}, the convergence of the sequence associated with \bmbptcon~is of course not guaranteed and may call for resummation methods.

\subsubsection{Pad\'e resummation}

The sequences defined through  Eqs.~\eqref{eq:projOP} and \eqref{eq:projvarOP} can be resummed using the well-known Pad\'e  scheme.  This resummation technique has proven to be successful in the context of standard MBPT~\cite{RothPade} and is briefly recalled in \ref{ap:pade}. 

\subsubsection{Eigenvector continuation}

Since no analytical property of the sequence is known, conventional resummation methods such as Padé cannot be applied with full confidence. Therefore, an alternative that does not rely on such a knowledge is highly desirable. Recently, a variational method coined as eigenvector continuation (EC)~\cite{Frame} was designed to treat physical systems whose Hamiltonian depends on a continuous control parameter that takes a specific value for the actual problem of interest. In the present context, the operator of interest is $\Omega (x)$ introduced in Eq.~\eqref{eq:split}. It continuously depends on the  control parameter $x$ scaling the residual interaction $\Omega_1$ and taking the value $x=1$ for the physical system of interest.

The rationale of the EC method relies on two principles:
\begin{enumerate}
	\item There exists a regime of the control parameter, e.g. $0 \leq x \leq x_e < 1 
	$ for which the system is easier to solve than for the physical value ($x=1$);
	\item When $x$ is varied back to $x=1$, the extremal eigenvectors of $\Omega (x)$
	only visit a low-dimensional sub-manifold of the 
	Hilbert space, i.e. the extremal eigenvectors trace out trajectories with a 
	significant displacement only in a few linearly-independent directions. 
\end{enumerate}
Presently, the first principle demands that, even if the problem of actual interest is not perturbative, it does become perturbative for small enough values of $x$.  Given that $\Omega (x)$ is expressed in a finite basis, the perturbative series indeed has a non-vanishing radius of convergence, i.e. there exists $x_e \leq 1$ such that the Taylor series of Eq.~\eqref{eq:exp} does converge for  $0 \leq x \leq x_e$. The smoothness of the problem ensures that the second principles applies, i.e. finding a low-dimensional manifold of eigenvectors in the well-behaved regime, they can be extrapolated to $x=1$ even when the perturbative expansion is not converging.  This extrapolation technique can in fact be understood as a sequence of analytic  continuations allowing to go beyond the radius of convergence of the perturbative expansion. 

In practice, the EC method consist of two successive steps:
\begin{enumerate}
	\item A low-dimensional manifold of $N_{EC}$ auxiliary states $\big \lbrace|\Psi^{[P]}_{n}(x_i)\rangle; 
	i=1,\cdots N_{EC} \big \rbrace$ 
	is obtained through BMBPT by computing the $P$-order eigenvectors of $\Omega (x_i)$ for a selection of $N_{EC}$ values  $0 \leq x \leq x_e$; 
	\item The targeted operator $\Omega=\Omega(1)$ is diagonalized in the low-dimensional 
	manifold obtained through step~1.  The auxiliary states being non-orthogonal, solving
 the secular equation requires to compute two $N_{EC}\times N_{EC}$ matrices,
 i.e., the grand potential $\langle \Psi^{[P]}_{n}(x_i) | \Omega | \Psi^{[P]}_{n}(x_j) \rangle$ and norm $\langle \Psi^{[P]}_{n}(x_i) | \Psi^{[P]}_{n}(x_j) \rangle$ kernels.
\end{enumerate}
Given Eq.~\eqref{eq:setaproxD},  one notices that
\begin{equation}
\begin{pmatrix}
|\Psi_{n}^{[P]}(x_1)\rangle
\\\vdots
\\|\Psi_{n}^{[P]}(x_{N_{EC}})\rangle
\end{pmatrix}=
\begin{pmatrix}
1       & x_1     & x_1^2   & \cdots  & x_1^P\\
1       & x_2     & x_2^2   & \cdots  & x_2^P\\
\vdots  & \vdots  & \vdots  & \ddots  & \vdots\\
1       & x_{N_{EC}}     & x_{N_{EC}}^2   & \cdots  & x_{N_{EC}}^P\\
\end{pmatrix}
\begin{pmatrix}
|\Phi_{n}^{(0)}\rangle
\\\vdots
\\|\Phi_{n}^{(P)}\rangle
\end{pmatrix} \ ,
\end{equation}
which implies that the secular equation can be equivalently written in terms of the set\footnote{The reader is reminded that these states should carry an extra subscript $P$ to underline the fact that they explicitly depend on the order $P$ at which one is working.}  $ \big\lbrace \ket{\Phi_{n}^{(p)}}; p=0, \cdots P \big \rbrace $ and does not actually depend on the choice of the $N_{EC}$ \(x_i\) values. Correspondingly, the effective dimensionality of the problem is in fact set by the chosen order $P$ and not by $N_{EC}$. Eventually, the $(P+1) \times (P+1)$ grand potential and norm matrices of practical interest are defined as
\begin{subequations} \label{matricessecularEq}
	\begin{eqnarray}
	\boldsymbol \Omega^{pq} &\equiv&
	\langle \Phi_{n}^{(p)} | \Omega | \Phi_{n}^{(q)} \rangle \ , \\
	\boldsymbol N^{pq} &\equiv&
	\langle \Phi_{n}^{(p)} | \Phi_{n}^{(q)} \rangle \ ,
	\end{eqnarray}
\end{subequations}
such that the former is Hermitian and the latter is symmetric positive definite. The generalized eigenvalue problem to solve reads as
\begin{equation}\label{eq:genEV}
\boldsymbol \Omega X = \mathcal E \boldsymbol NX \ .
\end{equation}
In practice, the EC method requires the sole knowledge of the perturbative state corrections $ \big\lbrace \ket{\Phi_{n}^{(p)}}, p=1, \cdots P 
\big \rbrace $ from which the grand potential and norm matrices can be computed. Still, instead of simply summing the corrections as is done in the Taylor expansion, the EC scheme adds a supplementary diagonalization step by solving the  Eq.~\eqref{eq:genEV}. 

In principle, the EC can be set up for any eigenstate characterized by the label $n$. Presently, the ground-state is targeted such that $n$ is set to $0$ in Eq.~\eqref{matricessecularEq}. Still, the associated secular equation  provides $P+1$ states $\ket{\Psi_{k \, EC}^{[P]}}$ among which the lowest one ($k=0$) is logically associated with the ground state. In addition, the other states ($k>0$) can be empirically compared to the $P$ lowest-lying excited states. Accessing excitation spectra via the EC technique deserves further investigation. In Sec.~\ref{sec:res} below, only results for the ground state are displayed and discussed. 

Having access to a new approximation to the eigenstates, one can compute any associated observable $O$ through the projective measure
\begin{equation}\label{eq:ECeval}
\mathcal O_{k \, EC}^{[P]} \equiv \Re \left\lbrace \bra{\Phi^{(0)}_{k}} O 
\ket{\Psi_{k \, EC}^{[P]}} \right\rbrace \ .
\end{equation}
While the present paper displays numerous BMBPT-based EC results, a preliminary highlight was already presented in Ref.~\cite{Demol:2019yjt} to disclose the merits and the potential of the approach.

\subsection{Reference results}

When working in a fixed subspace of manageable dimension defined by a subset of the unperturbed states $\{\ket{\Phi_{n}^{(0)}}\}$, it is possible to diagonalize the matrix representing a given operator, e.g. $\bra{\Phi_{p}^{(0)}} \Omega \ket{\Phi_{q}^{(0)}}$, to obtain its exact eigenvectors and eigenvalues in that subspace. With these eigenvectors at hand, any other observable can be computed. It corresponds to a (truncated) configuration interaction (CI) approach delivering reference results against which approximate methods implemented in the same subspace can be benchmarked.  

While CI calculations traditionally employ a many-body basis made out of $n$-particle/$n$-hole excited determinants, the present diagonalization is formulated within a subspace spanned by selected quasi-particle excitations of the Bogoliubov reference state. Consequently, the corresponding method is coined as (truncated) \emph{Bogoliubov configuration interaction} (BCI) and provides reference results for those obtained in the same subspace via BMBPT or via the resummation methods based on it discussed in \ref{resummationmethods}. Because the BCI method is subject to the same considerations as BMBPT regarding the handling of the average particle number, several variants will have to be considered in practice, i.e. \ciuncon, \cicon~or \ciap. Details of the BCI method are given in  \ref{sec:ED}. 

\subsection{Particle-number adjustment \label{sec:PNA}}

In Sec.~\ref{MBPTsetup}, \bmbptcon~was formally introduced to account for the contributions making up the average particle number at each order $P$. Adapting the average particle number carried by the reference state through the adjustment of the Lagrange parameter  $\lambda_P$ leads to the use of a $P$-dependent grand potential $\Omega_P$. In this way, one uses the freedom of choice of the reference  state to anticipate for the subsequent particle-number drift caused by BMBPT corrections such that the average particle number is eventually correct in the complete $P$-order\footnote{While the adjustment procedure is presently exemplified for \bmbptcon, it is equally valid for \bmbptcon-Pad\'e, \bmbptcon-EC or \cicon. In each case, the average particle number  is computed through the method of choice.}
\begin{equation}
\mathcal A^{[P]}_{0 P} = \text{A} \, .
\end{equation}

\begin{figure}[t]
	\centering
	\includegraphics[width=1.\textwidth]{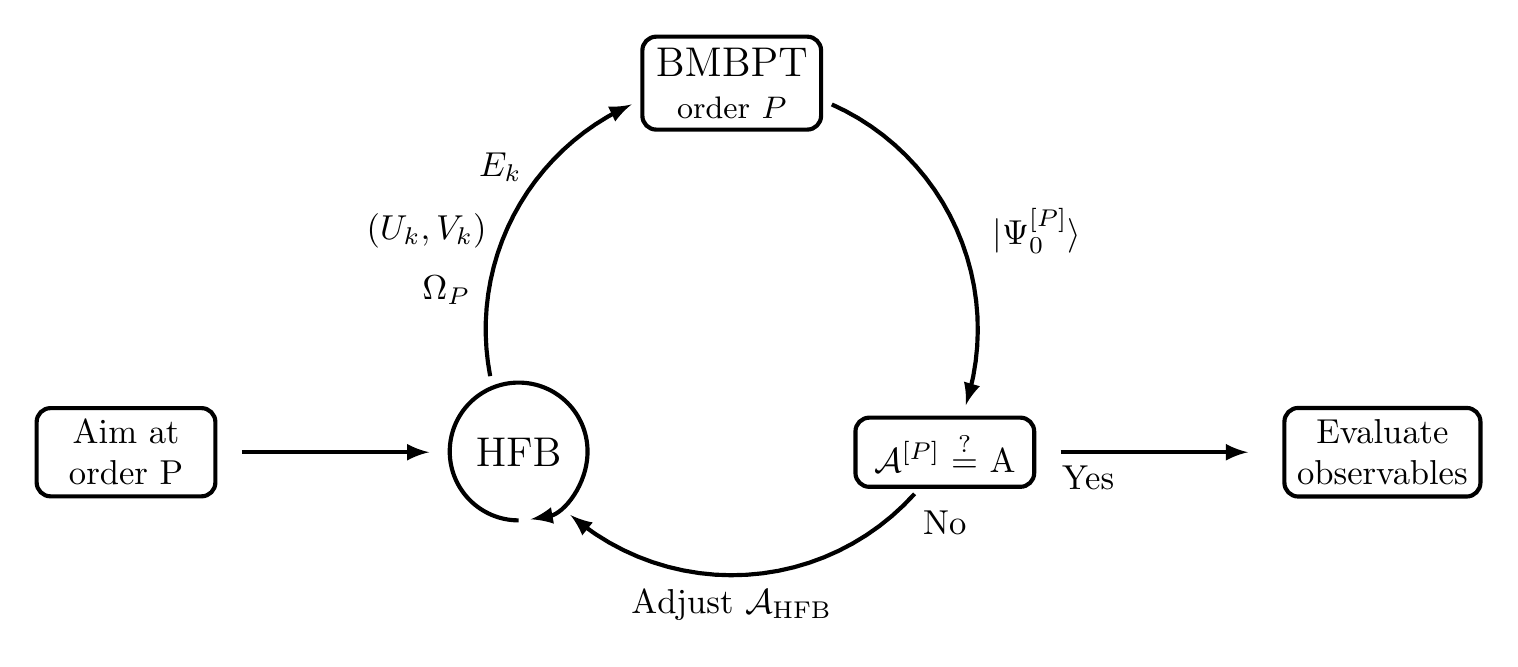}
	\caption{\small{\label{fig:PTUC}Schematic representation of the iterative 
	particle-number adjustment method in BMBPT for a specific order $P$.}}
\end{figure}

The intrinsically iterative character of the method is sketched in Fig.~\ref{fig:PTUC} for an arbitrary order $P$. One starts by solving the self-consistent HFB problem, described in \ref{ap:HFB}, imposing that the HFB vacuum contains A  particles on average. Next, e.g., BMBPT corrections are generated to compute $\mathcal A^{[P]}_{0 P}$. If $\mathcal A^{[P]}_{0 P} \neq $ A according to a certain accuracy measure, one recomputes the HFB reference state with a shifted average particle number that compensates for that difference, which sets $\lambda_P$ to a new value. Once again, BMBPT is solved to recompute $\mathcal A^{[P]}_{0 P}$ and the loop is in fact performed  until one reaches $\mathcal A^{[P]}_{0 P} = $ A. Once the iterative process is converged, other observables ($\Omega_P$, $H$ and $\Delta A$) are evaluated. This procedure needs to be repeated at each order $P$. 

\section{Results \label{sec:res}}

In this section, results obtained from BMBPT$^{(\circ / \bullet / \ast)}$ calculations and from resummation methods built on it are presented and systematically benchmarked against the corresponding BCI results. 

\subsection{Numerical implementation}\label{sec:num}

Having derived BMBPT equations along with the necessary ingredients to perform 
the particle-number adjustment in the previous chapter, the next step consists of 
numerically implementing this framework and testing its performance. Hence, an additional parallelized configuration-driven BMBPT code was implemented. 

The computations are performed using a realistic Hamiltonian derived from chiral EFT. The Hamiltonian contains a two-nucleon (NN) interaction derived at next-to-next-to-next-to leading order (N3LO) in the chiral expansion \cite{EntemMach} while the 3N interaction is omitted for simplicity. The bare chiral EFT interaction is not suited to a perturbative treatment due to its strong repulsive character at short distances. Hence, a SRG transformation is used to soften the interaction improving the convergence of the perturbative expansion~\cite{Bogner2007}. The SRG-softening however induces supplementary higher-body forces that are discarded beyond a certain particle rank\footnote{The normal-ordered two-body (NO2B) approximation discussed in Ref.~\cite{NOKB} can be used to take these higher-body forces partially into account as was done in Refs.~\cite{TichaiArthuis, Pierre}. In this application, all induced many-body forces of particle-rank three and higher are discarded, such that the Hamiltonian contains up to a two-body operator only.}. Therefore a trade-off between improved convergence and accuracy has to be made. In this application, the final SRG-flow parameter is $\alpha = 0.08$ fm$^4$. This value was shown to provide a convergent behaviour of the HF-based MBPT expansion in doubly closed-shell nuclei~\cite{TichaiMBPT}.  

The Hamiltonian is expressed in the one-body eigenbasis of the spherical harmonic 
oscillator Hamiltonian with frequency $\hbar \Omega = 20$ MeV 
using all single-particle states up to $e_{\text{max}} \equiv (2 n + l)_{\text{max}} 
= 4$. Realistic calculations typically make use of a model space characterized by
$e_{\text{max}}=12$ or 14 in order to reach convergence with respect to the basis 
set. Since the objective of this work is to investigate BMBPT 
at high orders, one is forced to perform these calculations in a small model space. 

The many-body configuration space, i.e. the subspace of $\mathcal F$ spanned by the eigenbasis $\{\ket{\Phi^{(0)}_n} = \ket{\Phi^{k_1\cdots k_i}}\}$ of $\Omega_0$ is also truncated. The used subspace consists of all single (two quasi-particles) and double (four quasi-particles) excitations as well as a portion of the triple (six quasi-particles) ones. The dominant triple configurations are incorporated via the use of importance truncation (IT) techniques \cite{IT} such that the configuration space is coined as $\mathcal F^{SD(T)}$. The basic idea behind IT is to estimate \textit{a priori} the importance of each state $\ket{\Phi^{k_1\cdots k_i}}$ using a computationally cheap method and discarding the irrelevant basis states. The IT measure\footnote{The IT measure used in this application is designed to minimize the loss of the associated third-order BMBPT correction.} employed is provided in Ref. \cite{IT}. The threshold below which states are discarded is denoted as $\kappa_{IT}$ and is taken equal to $10^{-6}$ in the present work. This value was shown to be sufficiently small to have no significant effect on the presented results. 

\subsection{Closed-shell system}

\begin{figure}[t]
	\centering
	\includegraphics[width=1.\textwidth]{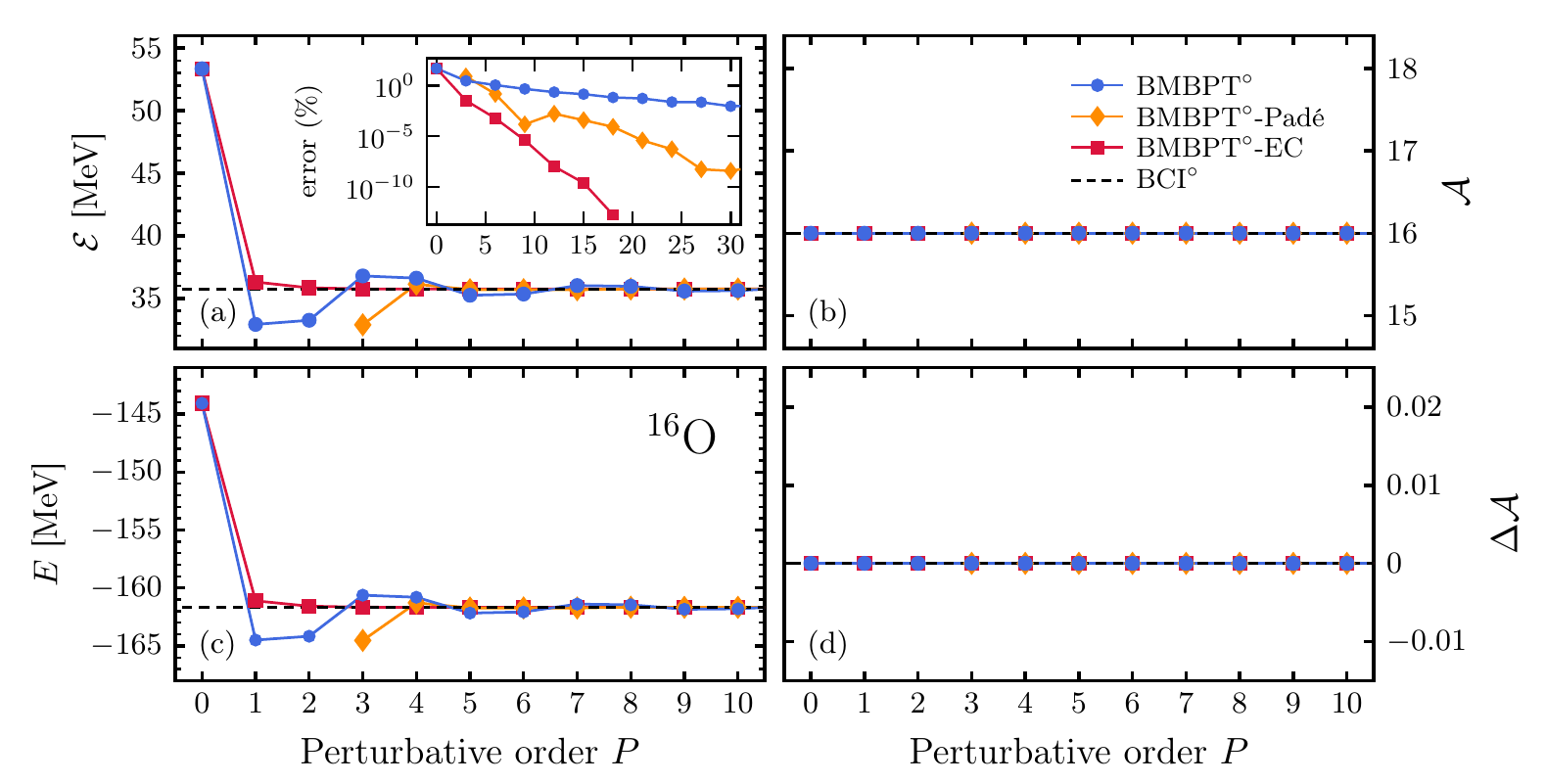}
	\caption{\label{img:unconAbsO16} $\nuc{16}{O}$ ground-state observables as a function of the perturbative order $P$ from \bmbptuncon (solid line/blue circles), \bmbptuncon-Pad\'e (solid line/yellow diamonds), \bmbptuncon-EC (solid line/red squares) and \ciuncon (dashed line). Panel (a): grand potential. Panel (b):  average particle number. Panel (c): energy. Panel (d): particle number dispersion. Panel (a) further includes an inset showing the relative error with respect to  \ciuncon. }
\end{figure}

As a first step, results obtained from the various methods introduced in Secs.~\ref{sec:obsexp}~and~\ref{sec:resum} are displayed in Fig.~\ref{img:unconAbsO16} for the ground state of the doubly closed-shell nucleus $\nuc{16}{O}$ as a function of the perturbative order $P$. The average particle number is trivially equal to $16$ independently of $P$ whereas the particle number dispersion vanishes systematically. These features are expected given that the Bogoliubov reference state reduces to a Slater determinant in doubly closed-shell nuclei such that BMBPT itself trivially reduces, in all of its unconstrained or constrained flavors, to the traditional particle-number-conserving MBPT. For the energy as well as for the grand potential, the \bmbptuncon~Taylor series converges to the \ciuncon~value, which itself reduces to standard CI in the present case. This result is consistent with the conclusions drawn in Ref. \cite{TichaiMBPT} where convergence properties of the HF-based MBPT were investigated.  

Given that \bmbptuncon~converges towards \ciuncon, the same is true for \bmbptuncon-Pad\'e and \bmbptuncon-EC resummation techniques. \bmbptuncon-EC present the advantage of converging from above thanks to its variational character. The associated convergence rates are compared in the inset for the grand potential by displaying the difference to the \ciuncon~result.  \bmbptuncon-EC converges the fastest, i.e. it already reaches $1\,\%$ accuracy for $P=2$. \bmbptuncon-Pad\'e only starts at third order and attains $1\,\%$ accuracy at order 4 while the strict Taylor expansion does so at $P=6$. 

\subsection{Open-shell system}

\subsubsection{Unconstrained BMBPT}

\begin{figure}[t!]
	\centering
	\includegraphics[width=1.\textwidth]{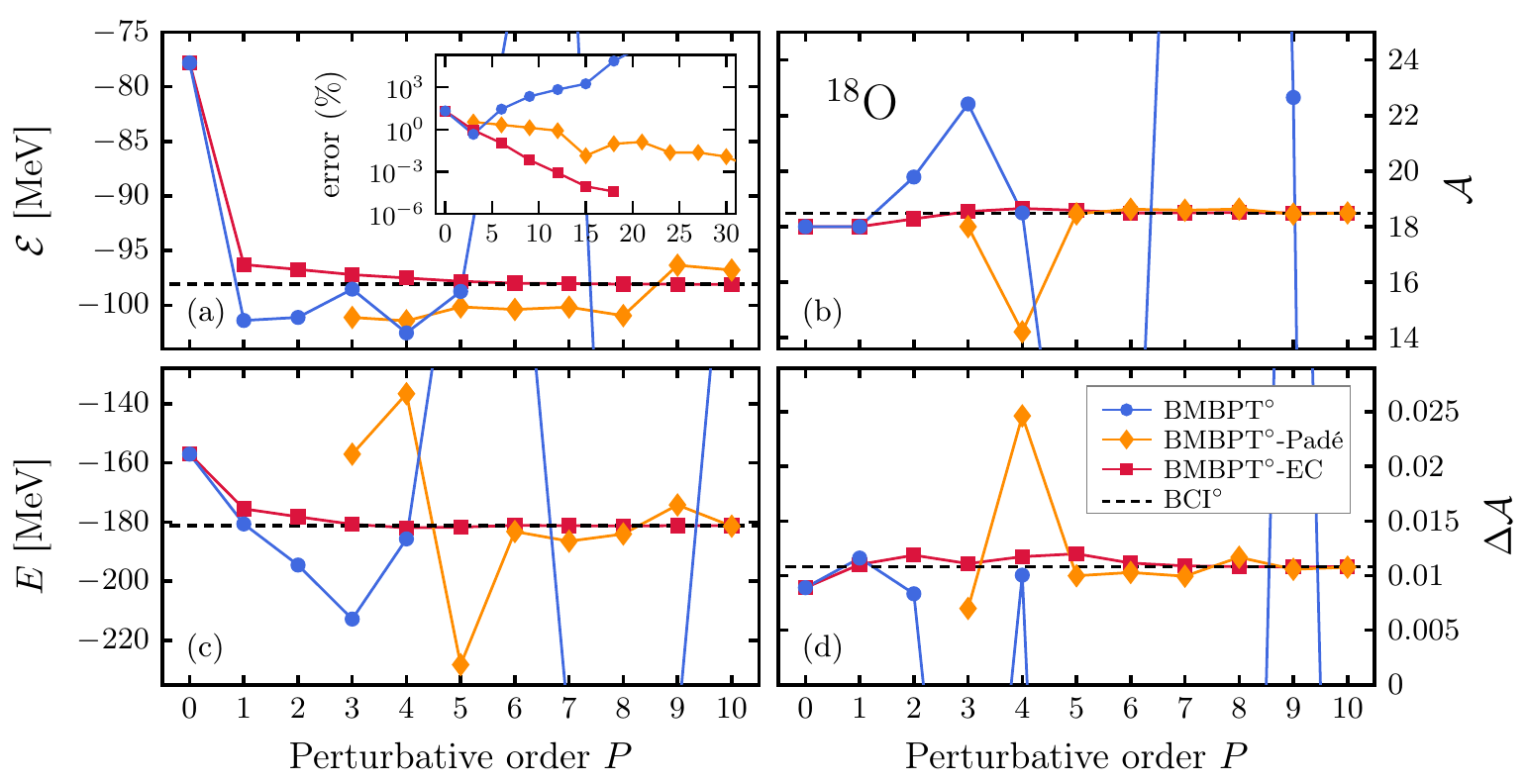}
	\caption{ \label{img:unconAbs} Same as Fig.~\ref{img:unconAbsO16} for $\nuc{18}{O}$. The particle number dispersion has been divided by $18^2$.}
\end{figure}

The study is now repeated for the ground state of the open-shell nucleus $\nuc{18}{O}$. 
First, \bmbptuncon~is investigated, i.e., the average particle number is constrained to the targeted value (A $=18$) only for the HFB reference state. Figure \ref{img:unconAbs} displays results in the same format as in Fig.~\ref{img:unconAbsO16}.  
The first striking finding is that the strict Taylor expansion diverges whereas \bmbptuncon-Pad\'e and \bmbptuncon-EC converge towards \ciuncon, although at a different rate and following different patterns.
In particular, the EC converges much faster and monotonically from above. 

The average particle number starts drifting at second order for \bmbptuncon~and eventually explodes. The excess of particles is reflected in the energy that displays a $15$ ($30$) MeV overbinding at second (third) order.  This feature is not observed for \bmbptuncon-EC that converges rapidly and variationally towards \ciuncon \footnote{Interestingly, the \ciuncon value slightly differs from the physical value $A=18$ given that the diagonalization takes place in a truncated subspace that is itself spanned by particle-number breaking basis states. Still, this shifted value does act as the reference for the approximate many-body methods implemented in the same subspace.}. \bmbptuncon-Pad\'e results are unreliable at low orders but converge rather quickly for $P\geq 5$ towards \ciuncon~as well. 

The particle-number dispersion is found to be different from zero in all cases\footnote{While the particle number dispersion is ensured to be positive when evaluated via the expectation value approach, it is not the case in the projective approach employed here. Additionally, while having a zero dispersion is a sufficient condition to ensure that the state is an eigenstate of $A$ when using the expectation value approach, it is only a necessary condition in the projective approach. In the projective approach, all moments $A^k$ must be equal to the number A$^k$ for the state under consideration to be an eigenstate of $A$.}. The finite dispersion delivered by \ciuncon~is again an artefact of the model-space truncation and would eventually go to zero in the full configuration space limit. In addition to being non-zero at low orders, the particle-number dispersion obtained from \bmbptuncon~quickly diverges with increasing orders. Contrarily, \bmbptuncon-Pad\'e and \bmbptuncon-EC resummation methods resolve the problem and converge towards the \ciuncon~value in the truncated model space. This convergence is once again faster for \bmbptuncon-EC than for \bmbptuncon-Pad\'e.

The convergence/divergence rates of the sequences are characterized in the inset.  \bmbptuncon, which in this unconstrained case is nothing but the consecutive partial sums of a single Taylor series, is shown to diverge exponentially. While both \bmbptuncon-Pad\'e and \bmbptuncon-EC converge, the latter does indeed do so at a faster (eventually exponential) rate.

\begin{figure}[t!]
	\centering
	\includegraphics[width=\textwidth]{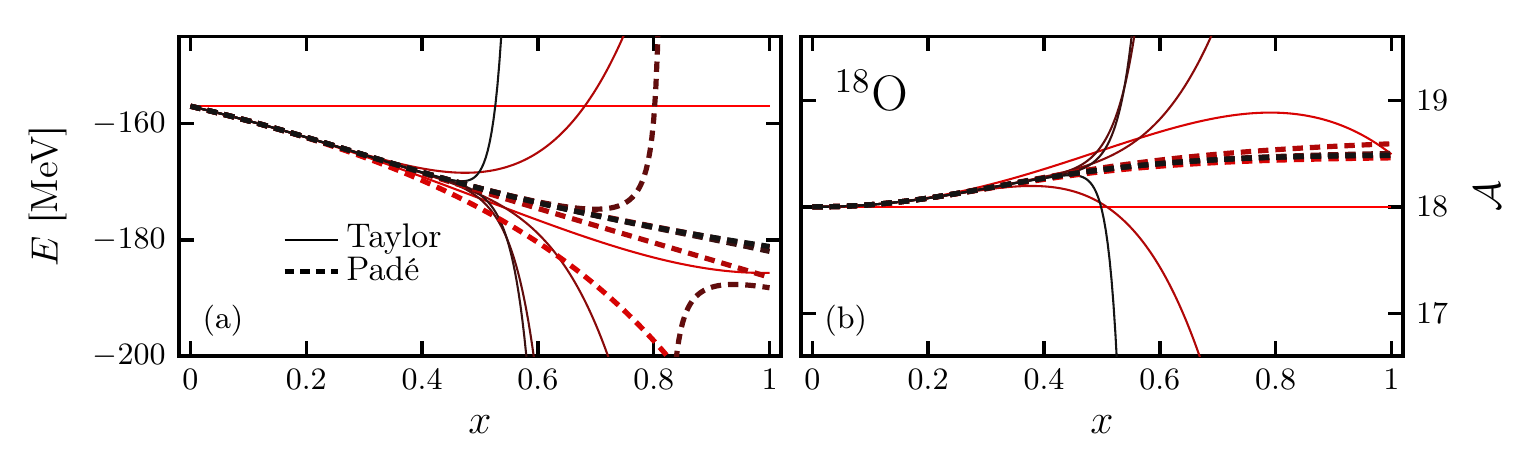}
	\caption{\label{img:TaylorPade}(color online) $\nuc{18}{O}$ ground-state (a) energy and (b) particle number obtained from \bmbptuncon (solid lines) and \bmbptuncon-Pad\'e (dashed lines) as a function of the expansion parameter $x$. In each panel, increasing perturbative orders ($P$=\{0,4,6,9,15,20,30\}) correspond to increasingly darker curves.}
\end{figure}

The divergence of the \bmbptuncon~Taylor series is now further investigated  by taking a closer look at its successive partial sums as a function of the expansion parameter $x$. In Fig.~\ref{img:TaylorPade}, the Taylor series truncated at various orders is depicted for the energy and the particle number. A divergence around $x=0.5$ is observed such that the physical point ($x=1$) is clearly outside the radius of convergence of the Taylor series. Since the consecutive partial sums are smooth within the domain of convergence, a resummation scheme seems well suited to recover the asymptotic value at $x=1$. Indeed, \bmbptuncon-Pad\'e does overcome the divergence\footnote{While one Pad\'e approximant to the energy does display a divergence for $x\approx0.8$, it can be attributed to a pole located too close to the real axis.} around $x=0.5$. 

While \bmbptuncon~is found to diverge systematically, resummation techniques can safely retrieve a converging sequence. However, at each finite order (beyond first order), and even at convergence, the average particle number is shown to differ from the physical value. Clearly, \bmbptcon,  \bmbptcon-Pad\'e, \bmbptcon-EC and \cicon~making use of the particle-number adjustment procedure formalized in Sec.~\ref{sec:BMBPT}  need now to be invoked to overcome this limitation. 

\subsubsection{HFB dependence}
\label{HFBdep}

\begin{figure}[t!]
	\centering
	\includegraphics[width=1.0\textwidth]{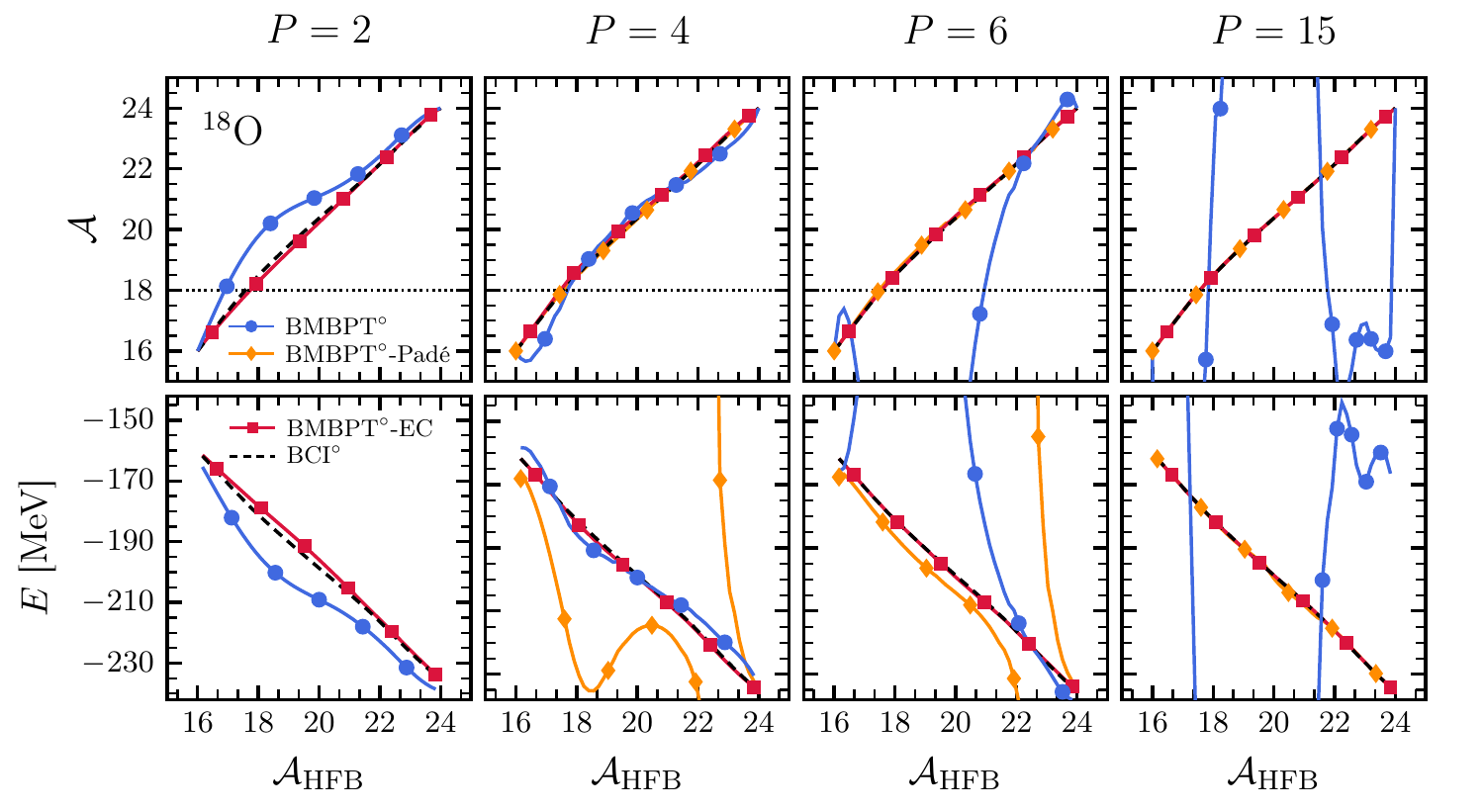}
	\caption{\label{img:AEvsHFB} $\nuc{18}{O}$ ground-state particle number (top panels) and energy (bottom panels) obtained from \bmbptuncon (solid line/blue circles), \bmbptuncon-Pad\'e (solid line/yellow diamonds), \bmbptuncon-EC (solid line/red squares) and \ciuncon (dashed line) as a function of the HFB average particle number at orders 2, 4, 6 and 15. }
\end{figure}

Since \bmbptcon~eventually relies on a careful adaptation of the average particle number $\mathcal A_{\text{HFB}}$ carried by the HFB reference state, it is worth examining the way observables computed at various orders in the unconstrained approaches depend on $\mathcal A_{\text{HFB}}$. Thus, Fig.~\ref{img:AEvsHFB} displays the energy and the particle number at orders 2, 4, 6 and 15 as a function of $\mathcal A_{\text{HFB}} \in [16,24]$.  Since $\mathcal A = 16$ and 24 correspond to closed-shell nuclei, no perturbative correction to the average particle number arises in these limit cases. 

One first observes that the BCI curve is a monotonic function of  $\mathcal A_{\text{HFB}}$ and crosses the physical value $\text{A}=18$ for $\mathcal A_{\text{HFB}} \approx 18$, i.e. the net shift brought by BCI correlations on top of the reference state are mild. Moving to BMBPT, $\mathcal A^{[P]}$ is also a monotonic function of $\mathcal A_{\text{HFB}}$ at low orders and follows quite closely the BCI curve. In contrast, the function becomes erratic and quickly oscillating for $P\geq 6$ such that no unique solution can be found for $\mathcal A^{[P]} = 18$ in the interval  $\mathcal A_{\text{HFB}} \in [16,24]$. Additionally, the function  $\mathcal A^{[P]}( A_{\text{HFB}})$ changes abruptly from one order to the next such that the HFB vacuum providing the correct average particle number at a given order $P$ does not relate in any transparent, i.e. continuous, way to the one found at order $P+1$. From the empirical standpoint, this confirms the inappropriate behaviour of the BMBPT Taylor series beyond the lowest orders. However, resumming the Taylor series through Pad\'e or EC does restore appropriate properties. The corresponding functions $\mathcal A^{[P]}( A_{\text{HFB}})$ are monotonic and quickly fall onto the BCI curve when increasing $P$.

Similar observations are made for the energy on the bottom panels\footnote{The patterns in the average particle number curve is reflected in an inverted way into the energy one. This relates to the fact that $E = \mathcal{E} + \lambda \mathcal A$ and that the grand potential is essentially independent on $A_{\text{HFB}}$.}. The Taylor expansion is well behaved at low orders but becomes erratic for $P>4$ and does not converge to the BCI curve. On the other hand, BMBPT-Pad\'e and BMBPT-EC do converge to BCI although poles contaminate the Pad\'e approximants at orders $4$ and $6$. 

Eventually, the Taylor expansion seems suitable to perform a particle-number adjustment at low orders, i.e. $P \lesssim 4$, since a unique HFB vacuum with $\mathcal A_{\text{HFB}} \approx 18$ is found as a solution to $\mathcal {A}^{[P]}(\mathcal A_{\text{HFB}}) = 18$. At higher orders, this is no longer possible and the use of a resummation method is mandatory to achieve a meaningful particle-number adjustment. 

\subsubsection{Constrained BMBPT}

\begin{figure}[t!]
	\centering
	\includegraphics[width=1.\textwidth]{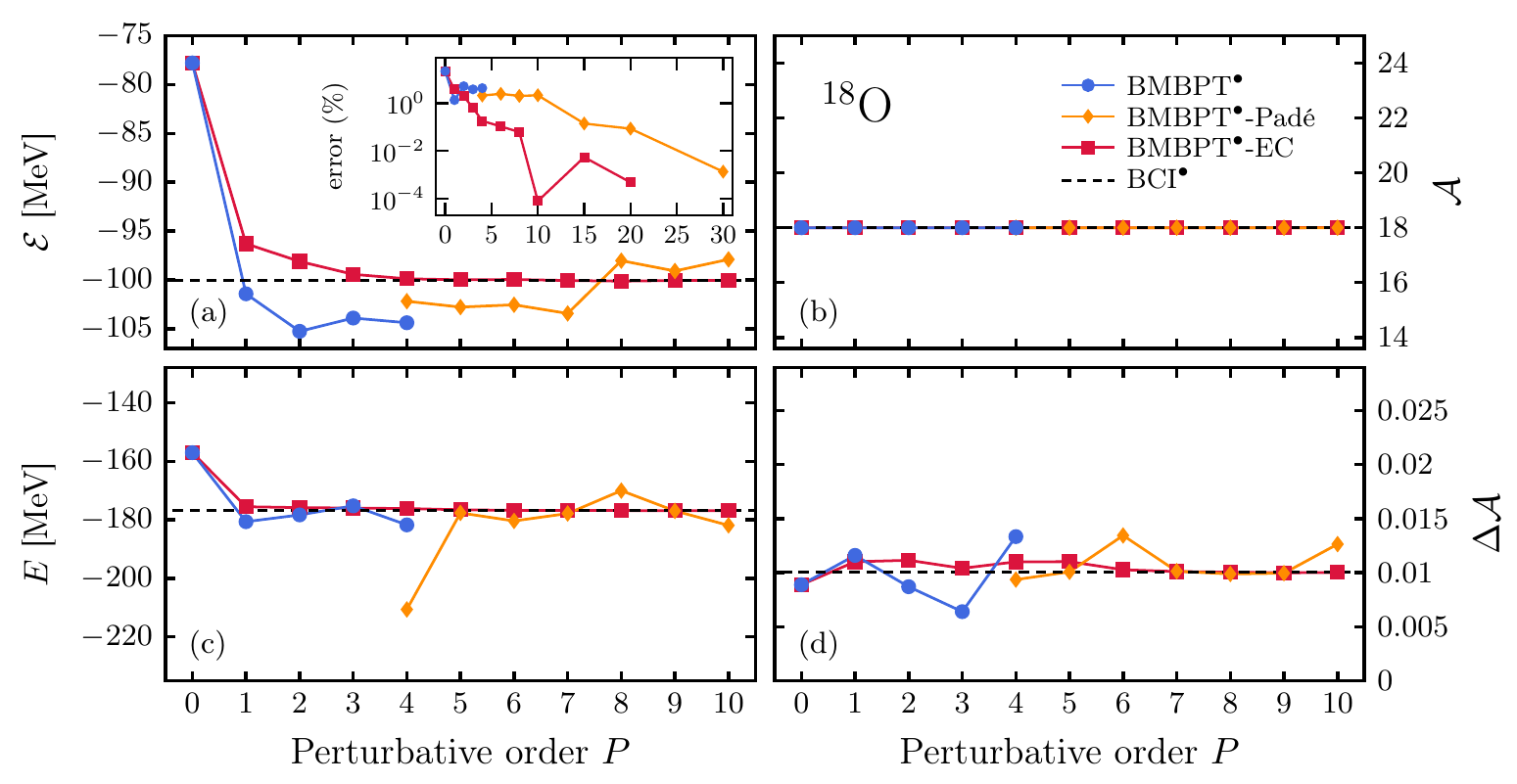}
	\caption{\small{\label{img:cons} Same as Fig.~\ref{img:unconAbs} for particle-number-constrained methods. Numerical parameters are $e_{\text{max}} = 4$, $\mathcal F^{SD(T)}$, $ \kappa_{\text{IT}}= 10^{-6}$.}}
\end{figure}

Results obtained with the particle-number adjustment are displayed in Fig.~\ref{img:cons}. In each case, the average particle number is evaluated consistently with the method of choice and is adjusted to the target value at each working order $P$ to better than $10^{-5}$. 

As seen from panel (b), the average particle number is indeed equal to 18 in all cases, meaning that the adjustment procedure succeeds in constraining the average particle number at each working order\footnote{The only exception is \bmbptcon-Pad\'e at third order for which no HFB vacuum resulting in $\mathcal A_{0}^{[3]} = 18$ could be found.}.  

As for the energy, \bmbptcon~performs well for the orders at which the constrained scheme is applicable, i.e. up to order 4. Resumming the series through \bmbptcon-Pad\'e and \bmbptcon-EC method provides sequences converging to the \cicon~limit. Knowing that \cicon~displays now the correct average particle number\footnote{Even though the average particle number is correct, the particle number dispersion associated with \cicon~is still (wrongly) different from zero due to the fact that the diagonalization operates in a truncated subspace.}, the associated reference energy differs slightly from the \ciuncon~one visible on Fig.~\ref{img:unconAbs} and can be considered as the optimal reference to reproduce. Taking a closer look at the grand potential through the inset of Fig. \ref{img:cons}(a), \bmbptcon-Pad\'e is shown to converge slowly, i.e. one must go to order 10 to reach an accuracy of about $1\,\%$. On the other hand, \bmbptcon-EC is quickly converging and is well below the $1\,\%$ accuracy at order 3.  

\subsubsection{\textit{A posteriori} correction}
\label{aposterioriconstraintmethod}

The particle-number adjustment procedure described in Sec.~\ref{sec:PNA} and employed in the previous section is computationally intensive. Consequently, a question of interest is whether or not it can be bypassed and safely replaced by an \textit{a posteriori} correction of the particle-number drift. This question is now addressed and results in the definition of a third flavour of BMBPT, denoted as \bmbptap, and of the resummation methods built on it. 

\begin{figure}[t]
	\centering
	\includegraphics[width=1.0\textwidth]{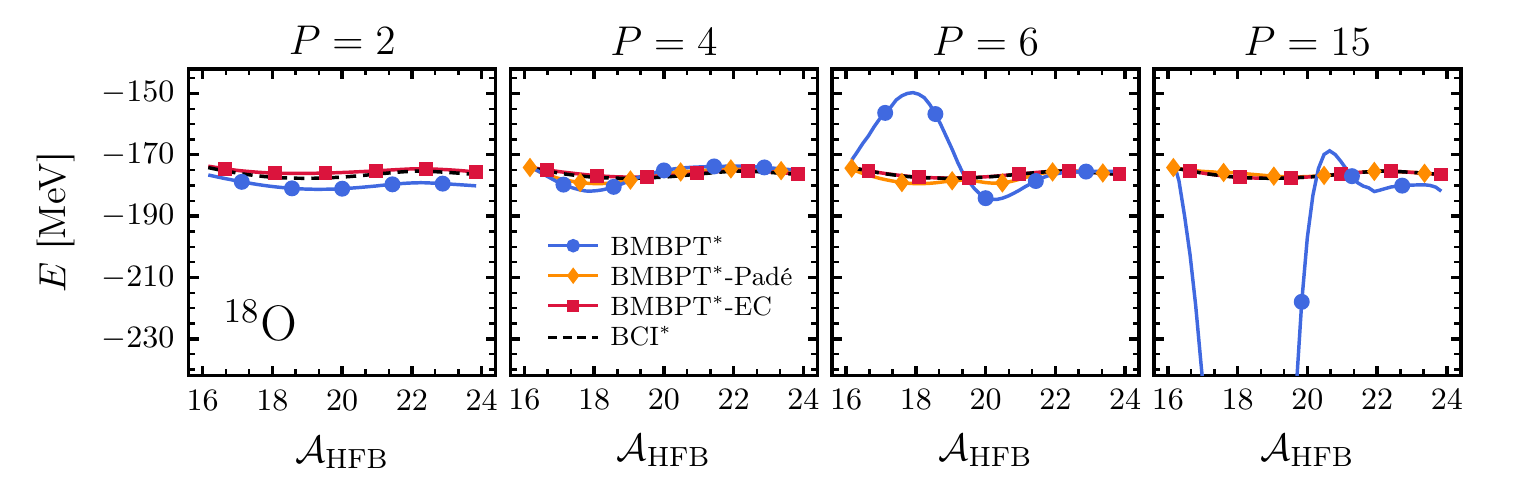}
	\caption{\label{img:OcvsHFB} \textit{A posteriori} corrected $\nuc{18}{O}$ ground-state energy obtained from \bmbptap (solid line/blue circles), \bmbptap-Pad\'e (solid line/yellow diamonds), \bmbptap-EC (solid line/red squares) and \ciap~(dashed line) as a function of the HFB average particle number at orders 2, 4, 6 and 15.}
\end{figure}

For any of the methods of interest, one has\footnote{In fact, 
Eq. \eqref{eq:evalP} is strictly 
true only for linear evaluation methods and therefore needs to be interpreted as an 
approximate equality when Pad\'e approximants are applied.}
\begin{equation}\label{eq:evalP}
E^{[P]} = \mathcal{E}^{[P]} + \lambda \mathcal A^{[P]} \ , 
\end{equation}
and
\begin{equation}
\left.E^{[P]}\right|_{\mathcal A^{[P]}+\delta \text{A}} \left.\approx 
E^{[P]}\right|_{\mathcal A^{[P]}} + \lambda \,  \delta\text{A} \ ,
\end{equation}
for a small variation of the average particle number. Substituting $\delta \text{A} 
\equiv \text{A} - \mathcal A^{[P]}$, i.e. the 
particle-number shift at order $P$, yields
\begin{equation}\label{eq:apost}
\left.E^{[P]}\right|_{\text{A}} \left.\approx E^{[P]}\right|_{\mathcal A^{[P]}} +
\lambda \left( \text{A} - \mathcal A^{[P]}\right)  =  \left. \mathcal E^{[P]} 
\right|_{\mathcal A^{[P]}}  + \lambda \text{A}\ .
\end{equation}
This can be used to correct for the drift $\text{A} - 
\mathcal A^{[P]}$ as long as it remains small compared to A. Applying this 
\textit{a posteriori} correction, there is no need to adjust the HFB vacuum at each
order $P$ such that the iterative particle-number adjustment procedure can be entirely circumvented.

The method is tested for $\nuc{18}{O} $ ground-state energy by computing $ E^{[P]} = \mathcal E^{[P]} + \lambda 18$ for a set of HFB reference states carrying $\mathcal A_{\text{HFB}}\in[16, 24]$. The obtained results are shown in Fig.~\ref{img:OcvsHFB}. First, one observes that the \ciap~curve is essentially flat, i.e. the energy is independent of $\mathcal A_{\text{HFB}}$ which is at variance with the linear dependence visible in Fig.~\ref{img:AEvsHFB}. It means that Eq.~\eqref{eq:apost} does indeed subtract from the energy the leading-order effect associated with the particle-number drift between the unperturbed vacuum and the correlated state. Second, the same behaviour is visible for  \bmbptap-Pad\'e and \bmbptap-EC as well as for \bmbptap~below $P=4$ due to the divergence of the series at higher orders. Again \bmbptap-EC performs best among all evaluation methods and  converges rapidly to the exact curve. 

\begin{figure}[t]
	\centering
	\includegraphics[width=.7\textwidth]{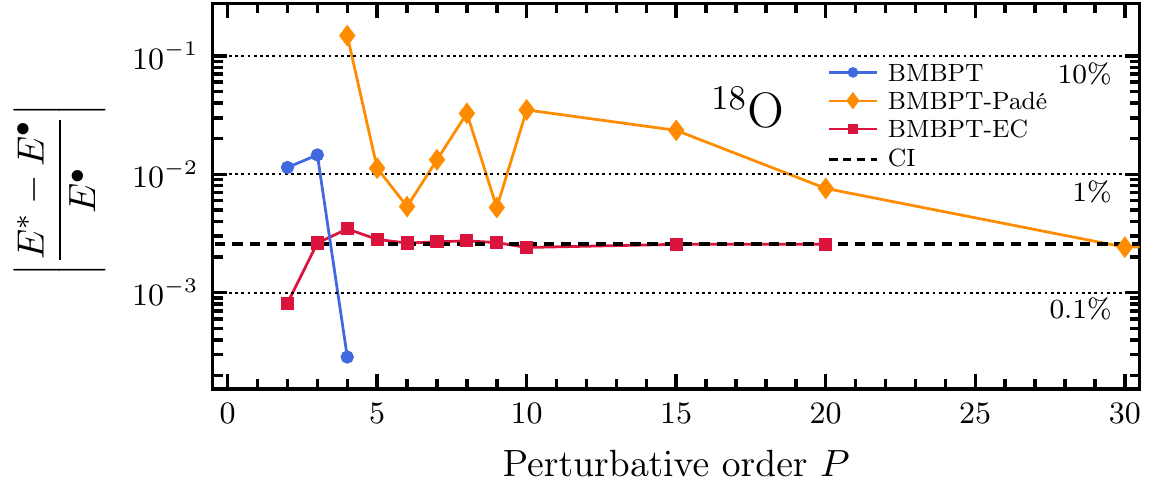}
	\caption{\label{img:apostvscons} Error of the \textit{a posteriori} corrected $\nuc{18}{O}$ ground-state energy obtained from \bmbptap (solid line/blue circles), \bmbptap-Pad\'e (solid line/yellow diamonds), \bmbptap-EC (solid line/red squares) and \ciap~(dashed line) with respect to results based on the self-consistent adjustment procedure as a function of the perturbative order $P$.}
\end{figure}

To gauge the accuracy of the \textit{a posteriori} correction, the error with respect to the results obtained through the self-consistent adjustment procedure is displayed in Fig. \ref{img:apostvscons} for all methods of interest. Up to order 4, the \textit{a posteriori} correction performs well for \bmbptap, providing a error below $1\,\%$\footnote{\bmbptap~results beyond fourth order are absent since the constrained calculations are not well defined in this regime; see Sec.~\ref{HFBdep}.}. For \bmbptap-Pad\'e, one needs to go beyond 5\ts{th} order to obtain accurate energies. Applying \bmbptap-EC, the \textit{a posteriori}  correction performs extremely well at all orders, eventually reaching a precision of about $0.2\,\%$, thus equating the one obtained for  \ciap~with respect to \cicon. 

\begin{figure}[t]
	\centering
	\includegraphics[width=.7\textwidth]{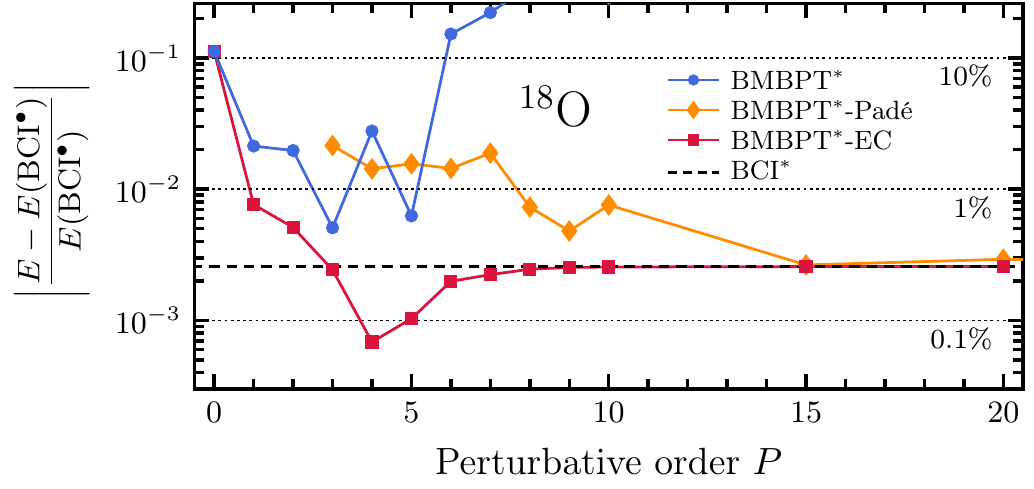}
	\caption{\label{img:apostvsexact} Error on the \textit{a posteriori} corrected $\nuc{18}{O}$ ground-state energy obtained from \bmbptap (solid line/blue circles), \bmbptap-Pad\'e (solid line/yellow diamonds), \bmbptap-EC (solid line/red squares) and \ciap~(dashed line) relative to  \cicon~results as a function of the perturbative order $P$.}
\end{figure}

The quality of \textit{a posteriori} corrected results can now be characterized by comparing them directly to \cicon~that constitutes the optimal reference method. As visible in Fig. \ref{img:apostvsexact}, \bmbptap~performs well up to order 4 where it delivers an error below $1\,\%$ but degrades quickly at higher orders. \bmbptap-Pad\'e converges towards \cicon, displaying a constant $2\,\%$ error for $P \in [1,7]$ before reaching the sub-percent accuracy for $P \geq 8$. Applying \bmbptap-EC, the \textit{a posteriori} correction performs extremely well, already reaching the sub-percent accuracy at first order before leveraging to the same $0.2\,\%$ error as \ciap~at higher orders.

One eventually concludes that the \textit{a posteriori} correction is  a cheap and accurate way to bypass the numerically costly particle-number adjustment procedure. While appropriate at low orders for the pure Taylor expansion, \bmbptap-EC delivers largely superior results and becomes mandatory at higher orders.

\subsection{Validation of low-order BMBPT calculations}

Realistic BMBPT calculations of mid-mass nuclei performed in large model spaces (e.g. $e_{\text{max}} = 12$, $\mathcal F^{SDT(Q)}$) will remain unattainable beyond $P=3$ for the years to come\footnote{So far, BMBPT has been implemented up to $P=2$, i.e. third order in the traditional counting~\cite{TichaiArthuis}.}. Moreover, the iterative particle-number-adjustment method being costly, employing \bmbptap~constitutes a preferable option for realistic calculations. 

\begin{figure}[t]
	\centering
	\includegraphics[width=1.\textwidth]{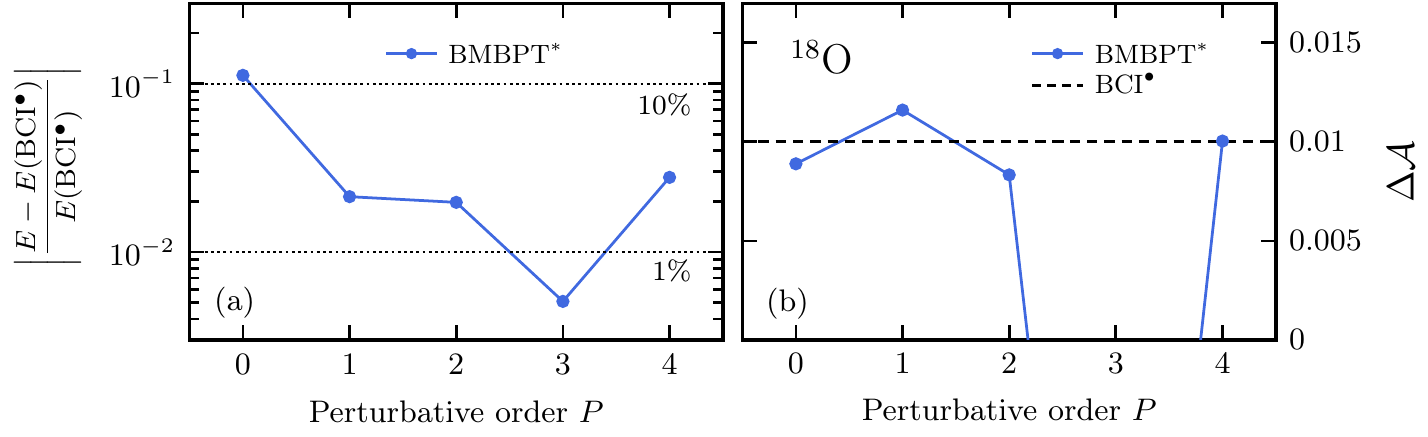}
	\caption{\label{img:lowo} Panel (a): Error on the \textit{a posteriori} corrected $\nuc{18}{O}$ ground-state energy obtained from \bmbptap (solid line/blue circles)  relative to  \cicon~results for the few lowest orders. Panel (b):  $\nuc{18}{O}$  ground-state particle-number dispersion obtained from \cicon and \bmbptap for $P \in [0,4]$.}
\end{figure}

Thus, Fig.~\ref{img:lowo} focuses on the accuracy achievable via low-order \bmbptap~calculations. Panel (a) demonstrates that a $2\,\%$ accuracy on the energy is typically reached at low orders compared to \cicon, which is similar to typical state-of-the-art non-perturbative methods and motivates the use of low-order \bmbptap~in future realistic calculations. However, panel (b) displaying the particle number dispersion illustrates that \bmbptap, while decent up to $P=2$, quickly behaves erratically. This feature underlines that, in spite of the adequate behaviour of the energy, low-order \bmbptap~results are contaminated by the breaking of $U(1)$ symmetry in a way that is not controlled. This eventually calls for the actual restoration of the symmetry via the recently designed particle-number projected BMBPT formalism~\cite{DuguetSignoracci}. While restoring the symmetry exactly, PBMBPT further incorporates additional static correlations into low-order calculations.

\section{Conclusions and outlook \label{sec:con}}

Convergence properties of the so-called Bogoliubov many-body perturbation theory (BMBPT), suited to the description of open-shell atomic nuclei have been investigated at length. 

The capacity of BMBPT to capture strong "static" correlations originates in the allowed breaking of $U(1)$ global gauge symmetry associated with the conservation of particle number. As a result, BMBPT was formulated as a perturbative expansion under the constraint that the particle number is correct {\it in average}. Subsequently, a recursive scheme was invoked to perform BMBPT calculations up to high orders to investigate the convergence behaviour of the associated Taylor series. Furthermore, resummation techniques from applied mathematics, i.e. well-known Pad\'e approximants and the novel eigenvector continuation  method~\cite{Frame}, were considered to evaluate observable beyond the strict Taylor expansion. 

Benchmark calculations of the $\nuc{16}{O}$ doubly closed-shell nucleus  were first shown to reduce to the standard particle-number conserving HF-MBPT and thus to generate a convergent Taylor series~\cite{TichaiMBPT}. The use of eigenvector continuation was found to accelerate the convergence rate substantially. In contrast, results for $\nuc{18}{O}$ demonstrated that the BMBPT Taylor expansion diverges in open-shell nuclei such that resummation techniques are required to transform the diverging sequence into a convergent one. 

In this context, the iterative adjustment procedure was shown to allow the constraint of the average particle number to the physical value at low orders in BMBPT, and thus to cure the energy from the contamination associated with the particle-number drift. Due to the divergence of the Taylor expansion, the use of resummation methods was however shown to be mandatory when going beyond order $4$. Furthermore, the \textit{a posteriori} correction on the average particle number used \textit{ad hoc} in Ref.~\cite{TichaiArthuis} was successfully benchmarked against the reference results obtained from constrained BMBPT.

Eventually, the key conclusions of the present investigation are that:
\begin{enumerate}
\item The full-fledged BMBPT under constraint can be bypassed via a low-cost {\it a posteriori} correction of the unconstrained/non-iterative BMBPT;
\item In spite of the diverging character of the Taylor expansion, low-order BMBPT predictions reproduce exact result within $2\,\%$;
\item The eigenvector continuation built from low-order BMBPT corrections presents a great potential to achieve high-accuracy results in realistic calculations whenever necessary. 
\end{enumerate}
Given that the results were obtained using a small, i.e. schematic, model space, the above conclusions cannot be naively extrapolated to realistic calculations. Still, one can genuinely hope that they will remain valid in much larger model spaces. Consequently, the next step will consist of investigating the above features in \emph{ab initio} calculations of open-shell nuclei performed in realistic model spaces. 

\section*{Acknowledgements}
We thank Robert Roth for providing us with nuclear matrix elements. This publication is based on work supported in part by Research Foundation Flanders (FWO, Belgium), by GOA/2015/010 (BOF KU Leuven), by
the framework of the Espace de Structure et de r\'eactions Nucl\'eaires Th\'eorique (ESNT) at CEA, and the Deutsche Forschungsgemeinschaft through contract SFB 1245.
Calculations were performed by using HPC resources from GENCI-TGCC (Contracts No. A005057392 and A007057392).


\appendix

\section{Hartree-Fock-Bogoliubov theory \label{ap:HFB}}

In this study, the Bogoliubov reference state is determined by solving the 
self-consistent Hartree-Fock-Bogoliubov (HFB) eigenvalue equation \cite{RingSchuck}. 
This mean-field equation is obtained by invoking Ritz' variational principle to minimise the total energy of the A-nucleon system under the constraint that 
$\bra{\Phi} A \ket{\Phi} = \text{A}$. One therefore requires that
\begin{equation}\label{eq:ritz}
\delta \left(\frac{\bra{\Phi} H \ket{\Phi}}{\overlap{\Phi}{\Phi}} - \lambda  
\frac{\bra{\Phi} A \ket{\Phi}}{\overlap{\Phi}{\Phi}} \right) = \delta 
\frac{\bra{\Phi} \Omega 
	\ket{\Phi}}{\overlap{\Phi}{\Phi}} = 0 \ , 
\end{equation}
where variations of the state $\ket{\tilde{\Phi}} = \ket{\Phi} + \ket{\delta \Phi}$ are 
considered to lay within the manifold of Bogoliubov states. 
Thouless' theorem~\cite{thouless} can be used to relate $\ket{\tilde{\Phi}}$ and 
$\ket{\Phi}$ explicitly. The theorem stipulates that two non-orthogonal Bogoliubov 
states can be connected through the non-unitary transformation
\begin{equation}\label{eq:thouless}
\ket{\tilde{\Phi}} = \overlap{\Phi}{\tilde{\Phi}} \ \text{exp} \left( \frac{1}{2} \sum_{k k'} 
Z_{kk'} \beta^{\dagger}_{k} \beta^{\dagger}_{k'}\right) \ket{\Phi} \ ,
\end{equation}
where 
\begin{equation}
Z\equiv \tilde{V}^{\ast}[\tilde{U}^{\ast}]^{-1} 
\end{equation}
is an antisymmetric matrix defined from the Bogoliubov transformation $(\tilde{U},\tilde{V})$
relating the quasi-particle operators of $\ket{\tilde{\Phi}}$ to those of $\ket{\Phi}$ 
introduced in Eq.~\eqref{eq:bt}.  Due to the anti-symmetry of $Z$, matrix elements $Z_{k_1k_2} $ with $k_1<k_2$ constitute the independent variational parameters.  Employing Eq.~\eqref{eq:thouless} and using that the norm in the denominator cancels disconnected terms, i.e. terms arising from contracting strings of quasi-particle 
operators that do not originate from the operator $\Omega$, one obtains
\begin{equation}
	\begin{aligned}\label{eq:varthou}
	\frac{\bra{\tilde{\Phi}} \Omega \ket{\tilde{\Phi}}}{\overlap{\tilde{\Phi}}{\tilde{\Phi}}}
	&= \Omega^{00} + \frac{1}{2} \sum_{k_1k_2} \left( 
	\Omega^{20}_{k_1k_2} Z^*_{k_1k_2} + \Omega^{02}_{k_1k_2} Z_{k_1k_2} \right) 
	 + \sum_{k_1k_2k_3} \Omega^{11}_{k_1k_2} Z^*_{k_1k_3} Z_{k_2k_3}\\ & \ \  +   
	 \frac{1}{8} \sum_{k_1k_2k_3k_4} \left( 
	\Omega^{40}_{k_1k_2k_3k_4} Z^*_{k_1k_2} Z^*_{k_3k_4} + 
	\Omega^{04}_{k_1k_2k_3k_4} Z_{k_1k_2} Z_{k_3k_4} + 2
	\Omega^{22}_{k_1k_2k_3k_4} Z^*_{k_1k_2} Z_{k_3k_4} \right)\ ,
	\end{aligned}
\end{equation}
where the expansion has been truncated to second order in $Z$. 
The variation of Eq.~\eqref{eq:varthou} with respect to $Z^*_{k_1k_2}$ evaluated at $Z=0$ is required to vanish for all $(k_1, k_2)$ pairs such that $k_1 < k_2$.  It provides the condition
\begin{equation}\label{eq:Omega20}
\frac{\partial}{\partial Z^*_{k_1k_2}} \left. \frac{\bra{\tilde{\Phi}} \Omega 
	\ket{\tilde{\Phi}}}{\overlap{\tilde{\Phi}}{\tilde{\Phi}}} \right|_{Z=0} = \frac{1}{2} 
\left(\Omega^{20}_{k_1k_2} - 
\Omega^{20}_{k_2k_1} \right)  
= \Omega^{20}_{k_1k_2} = 0 \ ,
\end{equation}
which is eventually valid for all $(k_1,k_2)$ due to the anti-symmetry of $\Omega^{20}$. 
Requiring the same for the variation with respect to $Z_{k_1k_2}$ gives the complementary equation
\begin{equation}\label{eq:Omega02}
\Omega^{02}_{k_1k_2} = 0 \ ,
\end{equation} 
for all $(k_1,k_2)$. 
One can conclude that $\Omega^{02}$ and $\Omega^{20}$ vanish in the 
quasi-particle basis associated to the HFB solution. 
However, Eqs.~\eqref{eq:Omega20} and~\eqref{eq:Omega02} do not 
constrain the form of the operator $\Omega^{11}$. 
Hence, this freedom allows one to  require a diagonal form of $\Omega^{11}$, i.e. $\Omega^{11}_{k_1k_2} \equiv \delta_{k_1k_2} E_{k_1}$. 
Combining this requirement with Eqs.~\eqref{eq:Omega20} and~\eqref{eq:Omega02} leads 
to obtaining the solution of the variational problem through the diagonalization of the matrix
\begin{equation}
\mathcal{\bar H} \equiv	
\begin{pmatrix}
\Omega^{11} & \Omega^{20}\\
\Omega^{02} & - \Omega^{11*}\\ 
\end{pmatrix} \ .
\end{equation}  
Expressing $\mathcal{\bar H}$ in the single-particle basis provides the Hartree-Fock-Bogoliubov Hamiltonian under the form
\begin{equation}
\mathcal H \equiv W \mathcal{\bar H} W^{\dagger} = 
\begin{pmatrix}
h- \lambda & \Delta\\
-\Delta^* & - (h-\lambda)^*\\
\end{pmatrix}  \ ,
\end{equation}
where the Hartree-Fock field $h$ and the Bogoliubov field $\Delta$ are defined through
\begin{subequations}
\begin{align}
	&h_{pq} \equiv t_{pq} + 
	\sum_{rs} \bar{v}_{psqr} \rho_{rs}  + \frac{1}{2} \sum_{rstu} \bar{w}_{prsqtu} 
	\Big( \rho_{us} \rho_{tr} + \frac{1}{2} \kappa^{*}_{rs}\kappa_{tu} \Big) \ ,\\
	&\Delta_{pq} \equiv \frac{1}{2} \sum 
	_{rs} \bar{v}_{pqrs} \kappa _{rs} + \frac{1}{2} \sum_{rstu} \bar{w}_{rpqstu} 
	\rho_{sr} \kappa_{tu} \ ,
\end{align}
\end{subequations}
with the one-body density matrices reading as
\begin{subequations}
\begin{align}
\rho_{pq}	& \equiv \frac{\bra{\Phi} c^{\dagger}_q c_p \ket{\Phi}}{\overlap{\Phi}{\Phi}} \ ,\\
\kappa _{pq} &  \equiv \frac{\bra{\Phi} c_q c_p \ket{\Phi}}{\overlap{\Phi}{\Phi}} \ .
\end{align}
\end{subequations}

Finding the variational minimum, i.e. the HFB reference state, amounts to diagonalizing $\mathcal H$, or equivalently to solving the eigenvalue equation
\begin{equation}\label{eq:HFBeigen}
\begin{pmatrix}
h- \lambda & \Delta\\
-\Delta^* & - (h-\lambda)^*\\
\end{pmatrix} 
\begin{pmatrix}
U_k\\
V_k\\
\end{pmatrix} 
= 
E_k
\begin{pmatrix}
U_k\\
V_k\\
\end{pmatrix} \ . 
\end{equation}
The eigenvectors $(U_k,V_k)$ determine the quasi-particle creation and annihilation operators $\{\beta_k^{\dagger},\beta_k\}$ through Eq.~\eqref{eq:bt} whereas the eigenvalues provide the quasi-particle energies $E_k$ entering Eq.~\ref{diagonebodypiece}. The eigenvalue problem of Eq.~\eqref{eq:HFBeigen} needs to be solved iteratively  until self-consistency is achieved given that $h$ and $\Delta$ depend on $U$ and $V$, i.e. on the eigenvectors. 

It is clear from Eq.~\eqref{eq:HFBeigen} that the HFB equation delivers $2n$ 
eigenvalues $E_k$ and eigenvectors $(U_k,V_k)$, where $n$ is equal to the dimension 
of the one-body Hilbert space  $\mathcal{H}_1$. 
In fact, these eigenstates appear in pairs: one with a positive quasi-particle energy $+ E_k$ and one as its negative counterpart with quasi-particle energy $- E_k$. 
The lowest-energy HFB solution is found by selecting the $n$ quasi-particle states with positive eigenvalues $E_k$ to build $\rho$ and $\kappa$, and thus $h$ and $\Delta$.

\section{Visited subspace of $\mathcal F$ \label{ap:subspace}}

\subsection{Many-body state
\label{ap:subspace_state}}

It is interesting to identify the unperturbed subspace of Fock space $\mathcal
F$ visited by the approximate ground state $\ket{\Psi^{[P]}_{0} (1)}$ defined 
through Eq.~\eqref{eq:stateOP} at a given order $P$. Unfolding Eq.~\eqref{eq:cnmp_rec}, state $\ket{\Phi_{0}^{(p)}}$ appears as a chain of excitations produced by acting $p$ times with the operator $\Omega_{1}$ on top 
of the Bogoliubov vacuum. The highest quasi-particle rank contained in 
$\ket{\Psi^{[P]}_{0} (1)}$ is therefore equal to $P$ times the quasi-particle rank 
of $\Omega_{1}$. Taking $\Omega_{1}$ to contain up to two-body terms as in the present applications, the first-order ground-state wave-function contains 
up to 4 quasi-particle excitations, i.e. its components lie in the so-called singles and 
doubles subspace $\mathcal F^{SD}$. Similarly, $\mathcal F^{SDTQ}$ is first reached 
at order $2$ and $\mathcal F^{SDTQPH}$ at order $3$. A more general statement is 
summarised in Tab.~\ref{t:subspacesC}.

\begin{table}[h]
	\centering
	\def\arraystretch{1.2}
	\begin{tabu}{|[2pt]c|[1.2pt]c|c|c|[2pt]}\tabucline[2pt]{-}
		\multirow{2}{*}{\makecell{Perturbative \\order $P$}} & 
		\multicolumn{3}{c|[2pt]}{rank of $\Omega_{1}$} 
		\\	\cline{2-4}
		& 2 & 3 & k 
		\\\tabucline[1.2pt]{-}
		0  & $\mathcal F^{0}$   & $\mathcal F^{0}$    & $\mathcal F^{0}$\\\hline
		1  & $\mathcal F^{SD}$  & $\mathcal F^{SDT}$   & $\mathcal
		F^{S\cdots2k}$\\\hline
		2  & $\mathcal F^{SDTQ}$  & $\mathcal F^{SDTQPH}$   & $\mathcal 
		F^{S\cdots4k}$\\\hline
		\vdots  & \vdots    & \vdots     & \vdots\\\hline
		$n$    & $\mathcal F^{S\cdots 4n}$   & $\mathcal F^{S\cdots6n}$     & 
		$\mathcal 
		F^{S\cdots2kn}$\\\tabucline[2pt]{-}
	\end{tabu}
	\caption{\label{t:subspacesC}Subspace of $\mathcal F$ contributing to 
		$\ket{\Psi^{[P]}_{0 P} (1)}$.}
	\label{tab:state}
\end{table}

Since $\Omega_{1}$ is presently represented in a fixed subspace of $\mathcal F$, the perturbative correction $\ket{\Phi_{n}^{(p)}}$ are complete only up to a certain order in BMBPT in practical applications. At higher 
orders, this truncation may have a significant impact since a large part
of Fock space which is in principle visited by the many-body state is ignored.

\subsection{Many-body observable \label{ap:subspace_obs}}

As for the wave function, it is interesting to 
investigate which subspace of $\mathcal F$ is visited through the Taylor-like 
evaluation of an observable at a given order $P$. Focusing on the ground state, Eqs.~\eqref{eq:cnmp_rec} and 
\eqref{eq:projOPex} indicate that the projective measure $\mathcal{O}_{0}^{[P]}(x)$ involves terms of the form
\begin{equation}
\sum_{q_0, q_1, \cdots, q_{p}} x^p \ 
\langle \Phi_{0}^{(0)}| O |\Phi_{q_0}^{(0)}\rangle
\langle \Phi_{q_0}^{(0)}| \Omega_{1} |\Phi_{q_1}^{(0)} \rangle
\cdots \langle\Phi_{q_{p-1}}^{(0)}| \Omega_{1}|\Phi_{q_{p}}^{(0)} \rangle 
C^{(0)}_{q_p0}
\ ,
\end{equation}
with $C^{(0)}_{q_p0} = \delta_{q_p0} $. Operators $O$ and $\Omega_{1}$ connect the unperturbed vacuum to itself 
through a chain of intermediate quasi-particle excitations. At order $P$, taking
$\Omega_{1}$ of rank $k$ and $O$ of rank $l$, the chain involves $2Pk+2l$
quasi-particle operators such that the visited space depends naturally on $l$ and 
$k$. Considering that $l\le k$ (which is the case in practical applications), the
most excited intermediate state has a quasi-particle rank equal to $2\left((P\mod2) 
l + \left\lfloor \frac P2 \right\rfloor k\right)$. This formula can be understood 
applying the following reasoning. At order $P$, half of the $\Omega_{1}$ 
operators act by exciting the vacuum while the other half is needed for the full 
de-excitation, giving rise to the term $\left\lfloor \frac P2 \right\rfloor k$. The 
remaining operator $O$ can induce $l$ additional excitations only if there is one 
$\Omega_{1}$ left for the subsequent de-excitation. This happens when $P$ 
is odd, hence the remaining term $(P\mod2) l$. A similar reasoning can be applied 
assuming $k < l \leq 2k$.  The cases $l \leq k$ and $k < l \leq 2k$ are 
summarised in Tab.~\ref{tab:subspaceTaylor}.

\begin{table}[h]
	\centering
	\def\arraystretch{1.2}
	\begin{tabu}{|[2pt]c|[1.2pt]c|c|[2pt]} \tabucline[2pt]{-}
		\makecell{Perturbative \\order $P$} &  $l \leq k$ & $k < l \leq 2k$ \\ 
		\tabucline[1.2pt]{-}
		0       & 0 & 0 \\\hline
		1       & $2 l$ & $2 k$ \\\hline
		2       & $2 k$ & $2 l$\\\hline
		3       & $2 (l+k)$& $2 (2k)$\\\hline
		4       & $2 (2k)$ & $2 (l+k)$\\\hline
		5       & $2 (l + 2k)$ & $2 (3k)$\\\hline
		6       & $2 (3k)$& $2 (l + 2k)$\\\hline
		\vdots  & \vdots & \vdots\\\hline
		$P = 2n$  	  & $2 (nk)$ & $2 (l+(n-1)k) $\\\hline
		$P = 2n+1$ 	  & $2 (l+nk) $ & $2(n+1)k$ \\ \tabucline[2pt]{-}
	\end{tabu}
	\caption{\label{tab:subspaceTaylor} Maximum quasi-particle rank of the subspace of 
		$\mathcal F$ contributing to $\mathcal{O}_{0}^{[P]} (1)$. $\Omega_{1}$ ($O$)
		is assumed to contain up to $k$-body ($l$-body) operators. Only the 
		cases $l \leq k$ and $k < l \leq 2k$ are represented.}
\end{table}

The variance of $O$ involves the square of that operator, which 
carries twice the rank. The subspace of $\mathcal{F}$ visited when evaluating the 
variance at a specific order is therefore larger. Table~\ref{tab:subspaceTaylor} can 
also be used in this case by simply doubling the rank of the considered operator. 

Due to the truncation of the active Fock space in practical applications, the evaluation of an observable is complete only up to a certain perturbative order $P$. 

\section{Pad\'e resummation \label{ap:pade}}

Given a function $\mathcal O(x)$, its $(M,N)$ Pad\'e approximant is defined
\cite{RothPade} as the unique rational function 
\begin{equation}
\mathcal O[M/N](x)=\
\frac{\sum_{i=1}^M a_ix^i}{1+\sum_{i=1}^Nb_ix^i} 
\end{equation}
satisfying
\begin{equation}
\left.\frac{\ud^k\mathcal O\left[M/N\right]}
{\ud x^k}\right|_{x=0} = \left.\frac{\ud^k\mathcal O}{\ud x^k}\right|_{x=0}
\ \forall\ 0 \le k \le M+N.
\end{equation}
The $(M,N)$ Pad\'e approximant of a function requires therefore the $M+N+1$ first
coefficients of its Taylor series. Denoting the Taylor series of $\mathcal O(x)$ as
\begin{equation}
\mathcal O(x) = \sum_{i=0}^\infty o_i \, x^i \ ,
\end{equation}
it can be obtained from the determinants of the 
$(N+1)\times(N+1)$ matrices containing the power-series coefficients $o_i$
\begin{equation}\label{eq:padeDet}
\mathcal O \left[M/N\right](x) \equiv
\dfrac{
	\begin{vmatrix}
	\sum_{k=0}^M o_kx^k & \sum_{k=0}^{M-1}o_kx^{k+1} & \cdots &\sum_{k=0}^{M-N} 
	o_kx^{k+N}\\
	o_{M+1} & o_{M}     & \cdots  & o_{M-N+1}   \\
	o_{M+2} & o_{M+1}   & \cdots  & o_{M-N+2} \\
	\vdots  & \vdots    & \ddots  & \vdots    \\
	o_{M+N} & o_{M+N-1} & \cdots  & o_{M}     \\
	\end{vmatrix}
}{
	\begin{vmatrix}
	1       & x       & \cdots  & x^N       \\
	o_{M+1} & o_{M}     & \cdots  & o_{M-N+1}   \\
	o_{M+2} & o_{M+1}   & \cdots  & o_{M-N+2} \\
	\vdots  & \vdots    & \ddots  & \vdots    \\
	o_{M+N} & o_{M+N-1} & \cdots  & o_{M}     \\
	\end{vmatrix}
} \ .
\end{equation}

The mathematical foundation of Pad\'e approximants relies 
on the conjecture formulated in Ref. \cite{Baker}. A simplified version 
reads: let the function $O(x)$ be a continuous function for $|x| \leq 1$, then there is an 
infinite subsequence of diagonal Pad\'e approximants $O[N/N] (x)$ that for $N \to 
\infty$ converges locally uniformly to $O(x)$ for $|x| \leq 1$.

Empirically, the use of a rational function instead of a Taylor series, i.e. a simple polynomial, 
is motivated by the fact that it can account for poles of the approximated function in the complex 
plane. These poles limit the convergence domain of the Taylor series while the 
Pad\'e approximant may converge on a larger domain, thus, being more flexible than classic Taylor series. Hence, Pad\'e 
approximants may converge to the asymptotic value of an observable, even though the 
partial sums of the Taylor series diverge. 

The Pad\'e resummation scheme is applied to the Taylor series of an observable 
$\mathcal O_{n}^{[P]} (x)$ given in Eq.~\eqref{eq:projOPex} for each order $P$. 
The  $(M,N)$ approximant depends only on the $M+N+1$ first coefficients of the power series 
and thus requires the knowledge of $\mathcal O_{n, M+N}^{[M+N]}(x)$.  The $P$-order 
Pad\'e approximant of an observable $O$ is eventually defined as
\begin{equation} \label{eq:pade}
\mathcal O^{[P]}_{n \, Pad\acute{e}}(x)\equiv \mathcal O_{n}^{[P]}
\bq{\floor{\tfrac P2} \big / {\ceil{\tfrac P2}} }(x).
\end{equation}
In fact, in order to resum only the dynamic correlations, i.e. the corrections appearing on top the HFB reference, the order-zero constant of the Taylor series is excluded when applying Eq.~\eqref{eq:pade}. The choice \((M,N)=(\floor{\tfrac P2},{\ceil{\tfrac P2}})\) is motivated by the
fact that \(M \sim N\). The $P$-order approximant for the dispersion is defined accordingly as
\begin{equation}
\Delta\mathcal O_{n \, Pad\acute{e}}^{[P]} (x) \equiv
{{\Delta \mathcal O}_{n }^{[P]}} \bq{\floor{\tfrac P2} \big / {\ceil{\tfrac P2}} }
(x)\ .
\end{equation}

\section{Bogoliubov configuration interaction \label{sec:ED}}

When working in a fixed configuration space, e.g. $\mathcal F^{SDT}$, it is possible to obtain eigenvectors of an operator expressed in that subspace of Fock space via an exact diagonalization. In the present case, the interest is to diagonalize the matrix $\bra{\Phi_{p}^{(0)}} \Omega \ket{\Phi_{q}^{(0)}}$ of the grand potential, i.e. to solve the eigenvalue problem
\begin{equation}
\sum_q \bra{\Phi_{p}^{(0)}} \Omega \ket{\Phi_{q}^{(0)}} 
\overlap{\Phi_{q}^{(0)}}{\Psi^{\EX}_n} = \mathcal E^{\EX}_n 
\overlap{\Phi_{p}^{(0)}}{\Psi^{\EX}_n} \ ,
\end{equation}
where  $\ket{\Psi^{\EX}_n}$ and $\mathcal E^{\EX}_n$ denote exact eigenstates and eigenvalues of the $\Omega$ matrix in the truncated space, respectively. 

Of course, the eigenvectors obtained in this way are not eigenstates of the full operator $\Omega$ due to the truncation effects induced by the restricted configuration space. Still, they provide pseudo-exact reference results for those obtained via BMBPT and associated resummation methods in the same subspace. Since $\Omega$ is Hermitian, the Lanczos algorithm~\cite{Lanczos,CullumWillloughby,Matrix} can be used to efficiently find extremal eigenvectors\footnote{Since in this application the ground state is presently targeted, one is interested in retrieving the eigenvector associated to the lowest eigenvalue of $\Omega$.}.  
 
Because the basis states making up the restricted configuration space are not eigenstates of the particle number operator, one must anticipate a shift of the average particle number carried by $\ket{\Psi^{\EX}_n}$ compared to the Bogoliubov vacuum $| \Phi \rangle$. Therefore two options arise\footnote{There exists a third option where the constraint is imposed via the order-$P$ evaluation of the average particle number on the basis of the Taylor,  Pad\'e or EC approach. Even though these hybrid methods might be interesting to investigate, only coherent methods using the same evaluation method for the 
constraint and the other observables are considered in this work.}:
\begin{enumerate}
\item One does not impose a constraint on the states generated via the diagonalization such that the particle-number constraint is only invoked for $| \Phi \rangle$, i.e. 
$P=0$. In analogy with \bmbptuncon, a subscript $0$ is added to indicate this choice, thus providing the states $\ket{\Psi^{\EX}_{n \, 0}}$ that do not carry the correct average particle number and the method is denoted as \ciuncon.
\item One requires  that $\ket{\Psi^{\EX}_n}$ carries the correct average particle number. The constraint is thus imposed on the output of the diagonalization by iteratively adjusting the reference state $| \Phi \rangle$ as described in Sec.~\ref{sec:PNA}. In this case, a subscript $\EX$ is added to indicate that the constraint is imposed on the exact eigenstate, 
leading to $\ket{\Psi^{\EX}_{n \, \EX}}$. In analogy with \bmbptcon, the method is coined as \cicon. 
\end{enumerate}

Once the eigenstates of the matrix are obtained, the associated observable $O$ can be evaluated in a projective fashion via
\begin{equation}
\mathcal O_{n \, (0/\EX)}^{\EX} \equiv \Re \left\lbrace  \bra{\Phi^{(0)}_{n \, (0/\EX)}} O 
\ket{\Psi^{\EX}_{n \, (0/\EX)}} \right\rbrace \ . 
\end{equation}

\section{Matrix elements in $\mathcal F^{SDT}$ basis \label{ap:ME} }

The working equations of BMBPT as well as the evaluation of an observable $O$ are 
formulated in terms of the matrix elements of the operators $\Omega_{1}$ and $O$ 
expressed in the unperturbed (configuration) basis $\ket{\Phi_{n}^{(0)}}$. 
In this appendix, the analytical expression of the matrix elements $\langle \Phi_n^{(0)} | O | 
\Phi_m^{(0)} \rangle$ of a generic operator $O$ in a generic basis made out of Bogoliubov 
states is derived.  Using Eq.~\eqref{eq:unperstates}, which identifies the zero-order states 
$|\Phi_m^{(0)} \rangle$ as quasi-particle excitations of the Bogoliubov vacuum, and 
writing $O$ according to Eq.~\eqref{eq:operator} yields
\begin{align}\label{eq:ME}
&\langle \Phi^{k_1\cdots k_a} | O | \Phi^{k_{a+1}\cdots k_{a+b}} \rangle
= \sum_{k = 0, 2, 4, 6,\cdots} \ \sum_{i+j=k} \ \langle \Phi^{k_1k_2\cdots 
	k_a} | O^{ij} | \Phi^{k_{a+1}\cdots	k_{a+b}} \rangle\\
&= \hspace{-2mm} \sum_{k = 0, 2, 4, 6,\cdots} \ \sum_{i+j=k} \dfrac{1}{i!j!} 
\hspace{-4mm} \sum_{\substack{k'_1, \cdots,k'_i \\ 
k'_{i+1},\cdots,k'_{i+j}}}\hspace{-4mm} O^{ij}_{k_1 \cdots k_i k_{i+1}	\cdots 
k_{i+j}} \langle \Phi |\beta_{k_{1}} \cdotss \beta_{k_{a}} \beta^{\dagger}_{k'_1} 
\cdotss \beta^{\dagger}_{k'_i} \beta_{k'_{i+j}} \cdotss \beta_{k'_{i+1}} 
\beta^{\dagger}_{k_{a+1}} \cdotss \beta^{\dagger}_{k_{a+b}} | \Phi \rangle \nonumber
\end{align}
Since $O^{ij}$ contains $i$ creators and $j$ annihilators, the contributions to Eq. 
\eqref{eq:ME} vanish as soon as $a-i \neq b-j$. 
However, satisfying the condition $a-i = b-j$ is not a sufficient condition to 
obtain a non-zero result. 
Quasi-particle indices of the bra and the ket state should be the same, up to a 
permutation, after acting with quasi-particle operators originating from $O^{ij}$.
Consequently,
\begin{equation}\label{eq:me}
\langle \Phi^{k_1\cdots k_a} | O | \Phi^{k_{a+1}\cdots k_{a+b}} \rangle = \sum_{k = 
0, 2, 4, 6,\cdots}\ \sum_{\substack{i+j=k\\i-j = a - b}} \ \sum_{\substack{l_1 < 
\cdots < l_i \\ \in \{ k_1,\cdots, k_a \}}} \ \sum_{\substack{l_{i+1} <\cdots 
<l_{i+j} \\ \in \{k_{a+1},\cdots, k_{a+b}\}}} \epsilon\ \delta\ O^{ij}_{l_1\cdots 
l_il_{i+1}\cdots l_{i+j}}
\end{equation}
where $\epsilon$ is equal to $\pm1$ depending on the permutation needed to have
the quasi-particle indices in the right order and $\delta$ is a shorthand
notation for $\delta_{ \{ k_1\cdotsm k_a \} \setminus \{ l_1\cdotsm l_i \} \ = \ 
\{k_{a+1} \cdotsm k_{a+b} \} \setminus \{l_{i+1}\cdotsm l_{i+j}\}}$. 
Essentially, all quasi-particle indices appearing only in one of the two states must 
be cancelled by the operator $O$. 
This compulsory set can then be enhanced by common indices of both states, if any. 

Using the above, the matrix of elements $\langle \Phi_{m}^{(0)}| O^{ij} | \Phi_{n}^{(0)} \rangle$ can be set up. In Eqs.~\eqref{eq:vtov}-\eqref{eq:ttot} below, the matrix of an operator $O$ containing a genuine three-body term and expressed in a basis of Bogoliubov states spanning $\mathcal F^{SDT}$ is provided. The quasi-particle labels of the Bogoliubov product states obey two consecutive ordering rules, i.e.
\begin{enumerate}
	\item Quasi-particle labels appearing in both the bra and the ket state are placed to the right;
	\item An arbitrarily-chosen sequential ordering $k_1<k_2<\cdots<k_n$ is imposed consistently within the subsets of common and unique quasi-particle labels. 
\end{enumerate}
With the latter specification, the indices are not meant to be naively exchanged. Still, the matrix elements corresponding to other label orderings are obtained from the same expressions by permuting the indices accordingly and by adding the sign given by the signature of the corresponding permutation. This table of matrix elements was used for the numerical implementation of the recursive BMBPT, although all terms associated with a genuine three-body operator were dropped in the applications discussed in the present paper. 

\subsection*{Vacuum to vacuum}
\begin{equation}\label{eq:vtov}
\bra{\Phi}{O}\ket{\Phi} = O^{00}
\end{equation}

\subsection*{Vacuum to single}
\begin{equation}\label{eq:vtos}
\bra{\Phi}{O}\ket{\Phi^{k_1k_2}} = O^{02}_{k_1k_2}
\end{equation}

\subsection*{Vacuum to double}
\begin{equation}\label{eq:vtod}
\bra{\Phi}{O}\ket{\Phi^{k_1k_2k_3k_4}} = O^{04}_{k_1k_2k_3k_4}
\end{equation}

\subsection*{Vacuum to triple}
\begin{equation}\label{eq:vtot}
\bra{\Phi}{O}\ket{\Phi^{k_1k_2k_3k_4k_5k_6}} = O^{06}_{k_1k_2k_3k_4k_5k_6}
\end{equation}

\subsection*{Single to single}
\begin{equation}\label{eq:stos}
\begin{aligned}
\bra{\Phi^{k_1k_2}}{O}\ket{\Phi^{k_3k_4}}  
= & \  O^{22}_{k_1k_2k_3k_4} + O^{11}_{k_1k_3} \delta_{k_2k_4} + O^{11}_{k_2k_4} \delta_{k_1k_3}
+ O^{00} \delta_{k_1k_3} \delta_{k_2k_4}
\end{aligned}
\end{equation}

\subsection*{Single to double}
\begin{equation}\label{eq:stod}
\begin{aligned}
\bra{\Phi^{k_1k_2}}{O}\ket{\Phi^{k_3k_4k_5k_6}}  
= & \ O^{24}_{k_1k_2k_3k_4k_5k_6} + O^{13}_{k_1k_3k_4k_5} \delta_{k_2k_6} + O^{13}_{k_2k_3k_4k_6} \delta_{k_1k_5} + O^{02}_{k_3k_4}\delta_{k_1k_5}\delta_{k_2k_6}
\end{aligned}
\end{equation}

\subsection*{Single to triple}
\begin{equation}\label{eq:stot}
\begin{aligned}
\bra{\Phi^{k_1k_2}}{O}\ket{\Phi^{k_3k_4k_5k_6k_7k_8}} 
= & \  O^{15}_{k_1k_3k_4k_5k_6k_7}\delta_{k_2k_8} + O^{15}_{k_1k_3k_4k_5k_6k_8}\delta_{k_1k_7} 
+ O^{04}_{k_3k_4k_5k_6}\delta_{k_1k_7}\delta_{k_2k_8}
\end{aligned}
\end{equation}

\subsection*{Double to double}
\begin{equation}\label{eq:dtod}
\begin{aligned}
\bra{\Phi^{k_1k_2k_3k_4}}{O}\ket{\Phi^{k_5k_6k_7k_8}} 
=& \ O^{33}_{k_1k_2k_3k_5k_6k_7} \delta_{k_4k_8} 
+ O^{33}_{k_1k_2k_4k_5k_6k_8} \delta_{k_3k_7}\\
&+ O^{33}_{k_1k_3k_4k_5k_7k_8} \delta_{k_2k_6}
+ O^{33}_{k_2k_3k_4k_6k_7k_8} \delta_{k_1k_5}\\
& + O^{22}_{k_1k_2k_5k_6} \delta_{k_3k_7} \delta_{k_4k_8}
+ O^{22}_{k_1k_3k_5k_7} \delta_{k_2k_6} \delta_{k_4k_8}\\
& + O^{22}_{k_2k_3k_6k_7} \delta_{k_1k_5} \delta_{k_4k_8}
+ O^{22}_{k_1k_4k_5k_8} \delta_{k_2k_6} \delta_{k_3k_7}\\
& + O^{22}_{k_2k_4k_6k_8} \delta_{k_1k_5} \delta_{k_3k_7}
+ O^{22}_{k_3k_4k_7k_8} \delta_{k_1k_5} \delta_{k_2k_6}\\
& + O^{11}_{k_1k_5} \delta_{k_2k_6} \delta_{k_3k_7} \delta_{k_4k_8}
+ O^{11}_{k_2k_6} \delta_{k_1k_5} \delta_{k_3k_7} \delta_{k_4k_8}\\
& + O^{11}_{k_3k_7} \delta_{k_1k_5} \delta_{k_2k_6} \delta_{k_4k_8}
+ O^{11}_{k_4k_8} \delta_{k_1k_5} \delta_{k_2k_6} \delta_{k_3k_7}\\
& + O^{00} \delta_{k_1k_5} \delta_{k_2k_6} \delta_{k_3k_7} \delta_{k_4k_8} 
\end{aligned}
\end{equation}

\subsection*{Double to triple}
\begin{equation}\label{eq:dtot}
\begin{aligned}
\bra{\Phi^{k_1k_2k_3k_4}}{O}\ket{\Phi^{k_5k_6k_7k_8k_9k_{10}}} 
= & \ O^{24}_{k_1k_2k_5k_6k_7k_8} \delta_{k_3k_9} \delta_{k_4k_{10}} 
+ O^{24}_{k_1k_3k_5k_6k_7k_9} \delta_{k_2k_8} \delta_{k_4k_{10}} \\
& + O^{24}_{k_2k_3k_5k_6k_8k_9} \delta_{k_1k_7} \delta_{k_4k_{10}} 
+ O^{24}_{k_1k_4k_5k_6k_7k_{10}} \delta_{k_2k_8} \delta_{k_3k_9} \\
& + O^{24}_{k_2k_4k_5k_6k_8k_{10}} \delta_{k_1k_7} \delta_{k_3k_9} 
+ O^{24}_{k_3k_4k_5k_6k_9k_{10}} \delta_{k_1k_7} \delta_{k_2k_8} \\
& + O^{13}_{k_1k_5k_6k_7}  \delta_{k_2k_8} \delta_{k_3k_9}  \delta_{k_4k_{10}}
+ O^{13}_{k_2k_5k_6k_8}  \delta_{k_1k_7}  \delta_{k_3k_9}  \delta_{k_4k_{10}}\\
& + O^{13}_{k_3k_5k_6k_9}  \delta_{k_1k_7}  \delta_{k_2k_8}  \delta_{k_4k_{10}}
+ O^{13}_{k_4k_5k_6k_{10}}  \delta_{k_1k_7}  \delta_{k_2k_8} \delta_{k_3k_9}\\
& + O^{02}_{k_5k_6}  \delta_{k_1k_7}  \delta_{k_2k_8} \delta_{k_3k_9}  \delta_{k_4k_{10}}  
\end{aligned}
\end{equation}

\subsection*{Triple to triple}
\begin{equation}\label{eq:ttot}
\begin{aligned}
\bra{\Phi^{k_1\cdots k_6}}{O}\ket{\Phi^{k_7\cdots k_{12}}}
= \ & O^{33}_{k_1k_2k_3k_7k_8k_9}\delta_{k_4k_{10}} \delta_{k_5k_{11}} \delta_{k_6k_{12}}
+ O^{33}_{k_1k_2k_4k_7k_8k_{10}}\delta_{k_3k_9} \delta_{k_5k_{11}} \delta_{k_6k_{12}}\\
& + O^{33}_{k_1k_3k_4k_7k_9k_{10}} \delta_{k_2k_8} \delta_{k_5k_{11}}\delta_{k_6k_{12}}
+ O^{33}_{k_2k_3k_4k_8k_9k_{10}}\delta_{k_1k_7} \delta_{k_5k_{11}} \delta_{k_6k_{12}}\\
& + O^{33}_{k_1k_2k_5k_7k_8k_{11}} \delta_{k_3k_9} \delta_{k_4k_{10}}\delta_{k_6k_{12}}
+ O^{33}_{k_1k_3k_5k_7k_9k_{11}} \delta_{k_2k_8}\delta_{k_4k_{10}} \delta_{k_6k_{12}} \\
& + O^{33}_{k_2k_3k_5k_8k_9k_{11}}\delta_{k_1k_7} \delta_{k_4k_{10}} \delta_{k_6k_{12}}
+ O^{33}_{k_1k_4k_5k_7k_{10}k_{11}} \delta_{k_2k_8}\delta_{k_3k_9} \delta_{k_6k_{12}}\\
& + O^{33}_{k_2k_4k_5k_8k_{10}k_{11}}\delta_{k_1k_7} \delta_{k_3k_9} \delta_{k_6k_{12}}
+ O^{33}_{k_3k_4k_5k_9k_{10}k_{11}}\delta_{k_1k_7} \delta_{k_2k_8} \delta_{k_6k_{12}}\\
& + O^{33}_{k_1k_2k_6k_7k_8k_{12}}\delta_{k_3k_9} \delta_{k_4k_{10}} \delta_{k_5k_{11}}
+ O^{33}_{k_1k_3k_6k_7k_9k_{12}}\delta_{k_2k_8}\delta_{k_4k_{10}} \delta_{k_5k_{11}}\\ 
& + O^{33}_{k_2k_3k_6k_8k_9k_{12}}\delta_{k_1k_7} \delta_{k_4k_{10}}\delta_{k_5k_{11}} 
+ O^{33}_{k_1k_4k_6k_7k_{10}k_{12}} \delta_{k_2k_8} \delta_{k_3k_9}\delta_{k_5k_{11}}\\
& + O^{33}_{k_2k_4k_6k_8k_{10}k_{12}}\delta_{k_1k_7}\delta_{k_3k_9}\delta_{k_5k_{11}} 
+ O^{33}_{k_3k_4k_6k_9k_{10}k_{12}}\delta_{k_1k_7} \delta_{k_2k_8} \delta_{k_5k_{11}}\\
& + O^{33}_{k_1k_5k_6k_7k_{11}k_{12}}\delta_{k_2k_8} \delta_{k_3k_9}\delta_{k_4k_{10}} 
+ O^{33}_{k_2k_5k_6k_8k_{11}k_{12}}\delta_{k_1k_7} \delta_{k_3k_9}\delta_{k_4k_{10}} \\
& + O^{33}_{k_3k_5k_6k_9k_{11}k_{12}}\delta_{k_1k_7} \delta_{k_2k_8}\delta_{k_4k_{10}}
+ O^{33}_{k_4k_5k_6k_{10}k_{11}k_{12}}\delta_{k_1k_7} \delta_{k_2k_8} \delta_{k_3k_9}\\
& + O^{22}_{k_1k_2k_7k_8}\delta_{k_3k_9} \delta_{k_4k_{10}} \delta_{k_5k_{11}} \delta_{k_6k_{12}}
+ O^{22}_{k_1k_3k_7k_9}\delta_{k_2k_8} \delta_{k_4k_{10}} \delta_{k_5k_{11}} \delta_{k_6k_{12}}\\
&  + O^{22}_{k_2k_3k_8k_9}\delta_{k_1k_7} \delta_{k_4k_{10}} \delta_{k_5k_{11}} \delta_{k_6k_{12}}
+ O^{22}_{k_1k_4k_7k_{10}} \delta_{k_2k_8} \delta_{k_3k_9} \delta_{k_5k_{11}} \delta_{k_6k_{12}}\\
&  + O^{22}_{k_2k_4k_7k_{10}}\delta_{k_1k_7} \delta_{k_3k_9}\delta_{k_5k_{11}} \delta_{k_6k_{12}}
+ O^{22}_{k_3k_4k_9k_{10}}\delta_{k_1k_7} \delta_{k_2k_8} \delta_{k_5k_{11}} \delta_{k_6k_{12}}\\
&+ O^{22}_{k_1k_5k_7k_{11}}\delta_{k_2k_8} \delta_{k_3k_9} \delta_{k_4k_{10}}  \delta_{k_6k_{12}}
+ O^{22}_{k_2k_5k_8k_{11}}\delta_{k_1k_7} \delta_{k_3k_9} \delta_{k_4k_{10}}  \delta_{k_6k_{12}}\\
&  + O^{22}_{k_3k_5k_9k_{11}} \delta_{k_1k_7} \delta_{k_2k_8} \delta_{k_4k_{10}}  \delta_{k_6k_{12}}
+ O^{22}_{k_4k_5k_{10}k_{11}}\delta_{k_1k_7} \delta_{k_2k_8} \delta_{k_3k_9} \delta_{k_6k_{12}}\\
&  + O^{22}_{k_1k_6k_7k_{12}} \delta_{k_2k_8} \delta_{k_3k_9} \delta_{k_4k_{10}} \delta_{k_5k_{11}} 
 + O^{22}_{k_2k_6k_8k_{12}}\delta_{k_1k_7} \delta_{k_3k_9} \delta_{k_4k_{10}} \delta_{k_5k_{11}} \\
& + O^{22}_{k_3k_6k_9k_{12}}\delta_{k_1k_7} \delta_{k_2k_8} \delta_{k_4k_{10}} \delta_{k_5k_{11}}
+ O^{22}_{k_4k_6k_{10}k_{12}}\delta_{k_1k_7} \delta_{k_2k_8} \delta_{k_3k_9}\delta_{k_5k_{11}} \\
&  + O^{22}_{k_5k_6k_{11}k_{12}}\delta_{k_1k_7} \delta_{k_2k_8} \delta_{k_3k_9} \delta_{k_4k_{10}}\\
& + O^{11}_{k_1k_7} \delta_{k_2k_8} \delta_{k_3k_9} \delta_{k_4k_{10}} \delta_{k_5k_{11}} \delta_{k_6k_{12}}
+ O^{11}_{k_2k_8} \delta_{k_1k_7} \delta_{k_3k_9} \delta_{k_4k_{10}} \delta_{k_5k_{11}} \delta_{k_6k_{12}}\\
& + O^{11}_{k_3k_9} \delta_{k_1k_7} \delta_{k_2k_8} \delta_{k_4k_{10}} \delta_{k_5k_{11}} \delta_{k_6k_{12}}
+ O^{11}_{k_4k_{10}} \delta_{k_1k_7} \delta_{k_2k_8} \delta_{k_3k_9} \delta_{k_5k_{11}} \delta_{k_6k_{12}}\\
& + O^{11}_{k_5k_{11}} \delta_{k_1k_7} \delta_{k_2k_8} \delta_{k_3k_9} \delta_{k_4k_{10}} \delta_{k_6k_{12}}
+ O^{11}_{k_6k_{12}} \delta_{k_1k_7} \delta_{k_2k_8} \delta_{k_3k_9} \delta_{k_4k_{10}} \delta_{k_5k_{11}}\\
& + O^{00} \delta_{k_1k_7} \delta_{k_2k_8} \delta_{k_3k_9} \delta_{k_4k_{10}} \delta_{k_5k_{11}} \delta_{k_6k_{12}}
\end{aligned}
\end{equation}

\bibliography{bibliography}

\end{document}